\documentclass[conference]{IEEEtran}
\ifCLASSINFOpdf
\else
\fi
\usepackage{amsmath,amsfonts}
\usepackage{graphicx}
\usepackage{textcomp}
\usepackage{bmpsize}
\usepackage{url}
\usepackage{lipsum}
\usepackage{float}
\usepackage{multicol}
\usepackage{makecell}
\usepackage{algorithm}
\usepackage{algpseudocode}
\usepackage{enumitem}
\usepackage[dvipsnames]{xcolor}
\usepackage{comment}

\usepackage{tikz}
\usetikzlibrary{patterns}

\usepackage{colortbl}

\usepackage{fancyhdr}
\usepackage{diagbox}
\usepackage{multirow}

\usepackage{makecell}
\usepackage{forest}
\usepackage{longtable}
\usepackage{tcolorbox}
\usepackage{pgf-pie}
\usepackage[caption=false]{subfig}
\usepackage{datapie}
\usepackage[]{mdframed}
\usepackage{graphicx}
\usepackage{booktabs}
\usepackage{multirow}
\usepackage{cleveref}

\crefformat{section}{\S#2#1#3} % see manual of cleveref, section 8.2.1
\crefformat{subsection}{\S#2#1#3}
\crefformat{subsubsection}{\S#2#1#3}

\makeatletter

\tikzstyle{chart}=[
    legend label/.style={font={\scriptsize},anchor=west,align=left},
    legend box/.style={rectangle, draw, minimum size=5pt},
    axis/.style={black,semithick,->},
    axis label/.style={anchor=east,font={\tiny}},
]

\tikzstyle{bar chart}=[
    chart,
    bar width/.code={
        \pgfmathparse{##1/2}
        \global\let\bar@w\pgfmathresult
    },
    bar/.style={very thick, draw=white},
    bar label/.style={font={\bf\small},anchor=north},
    bar value/.style={font={\footnotesize}},
    bar width=.75,
]

\tikzstyle{pie chart}=[
    chart,
    slice/.style={line cap=round, line join=round, very thick,draw=white},
    pie title/.style={font={\bf}},
    slice type/.style 2 args={
        ##1/.style={fill=##2},
        values of ##1/.style={}
    }
]

\pgfdeclarelayer{background}
\pgfdeclarelayer{foreground}
\pgfsetlayers{background,main,foreground}

\newcommand{\pie}[3][]{
    \begin{scope}[#1]
    \pgfmathsetmacro{\curA}{90}
    \pgfmathsetmacro{\r}{1}
    \def\c{(0,0)}
    \node[pie title] at (90:1.3) {#2};
    \foreach \v/\s in{#3}{
        \pgfmathsetmacro{\deltaA}{\v/100*360}
        \pgfmathsetmacro{\nextA}{\curA + \deltaA}
        \pgfmathsetmacro{\midA}{(\curA+\nextA)/2}

        \path[slice,\s] \c
            -- +(\curA:\r)
            arc (\curA:\nextA:\r)
            -- cycle;
        \pgfmathsetmacro{\d}{max((\deltaA * -(.5/50) + 1) , .5)}

        \begin{pgfonlayer}{foreground}
        \path \c -- node[pos=\d,pie values,values of \s]{$\v\%$} +(\midA:\r);
        \end{pgfonlayer}

        \global\let\curA\nextA
    }
    \end{scope}
}

\usepackage{tikz}
\usepackage{amsmath}

\newcommand*\circled[1]{\tikz[baseline=(char.base)]{
            \node[shape=circle,draw,inner sep=1.5pt] (char) {#1};}}

\usepackage{filecontents}

\definecolor{mygray}{gray}{0.93}
\definecolor{mypink}{RGB}{250, 187, 187}
\definecolor{myellow}{RGB}{255, 153, 0}
\definecolor{myred}{RGB}{255, 93, 93}
\definecolor{mygreen}{RGB}{111, 255, 111}
\definecolor{myazure}{RGB}{224, 255, 255}

\makeatletter
\newcommand\notsotiny{\@setfontsize\notsotiny\@vipt\@viipt}
\makeatother

\forestset{
  L1/.style={fill=white,},
  L2/.style={fill=myred,edge={myred,line width=2pt}},
  L3/.style={fill=yellow,edge={yellow,line width=2pt}},
  L4/.style={fill=mygreen,edge={mygreen,line width=2pt}},
}

\begin{document}
%\title{Introducing Perturb-ability Score (PS) to Enhance Robustness Against Problem-Space Evasion Adversarial Attacks on Flow-based ML-NIDS}

\title{A Novel Perturb-ability Score to Mitigate Evasion Adversarial Attacks on Flow-Based ML-NIDS}

%\title{Introducing Perturb-ability Score to Metigate Flow-Based ML-NIDS Against Problem-Space Evasion Adversarial Attacks}

% affiliations'

\author{\IEEEauthorblockN{Mohamed elShehaby}
	\IEEEauthorblockA{Carleton Univeristy\\
		MohamedelShehaby@cmail.carleton.ca}
	\and
	\IEEEauthorblockN{Ashraf Matrawy}
	\IEEEauthorblockA{Carleton Univeristy\\
		Ashraf.Matrawy@carleton.ca}
	}


%Telephone: (800) 555--1212, Fax: (888) 555--1212}


\IEEEoverridecommandlockouts
\makeatletter\def\@IEEEpubidpullup{6.5\baselineskip}\makeatother

\maketitle

\begin{abstract}
As network security threats evolve, safeguarding flow-based Machine Learning (ML)-based Network Intrusion Detection Systems (NIDS) from evasion adversarial attacks is crucial. This paper introduces the notion of {\em feature perturb-ability} and presents a novel {\em Perturb-ability Score (PS)}, which quantifies how susceptible NIDS features are to manipulation in the problem-space by an attacker. PS thereby identifies features structurally resistant to evasion attacks in flow-based ML-NIDS due to the semantics of network traffic fields, as these features are constrained by domain-specific limitations and correlations. Consequently, attempts to manipulate such features would likely either compromise the attack's malicious functionality, render the traffic invalid for processing, or potentially both outcomes simultaneously.

We introduce and demonstrate the effectiveness of our PS-enabled defenses, PS-guided feature selection and PS-guided feature masking, in enhancing flow-based NIDS resilience. Experimental results across various ML-based NIDS models and public datasets show that discarding or masking highly manipulatable features (high-PS features) can maintain solid detection performance while significantly reducing vulnerability to evasion adversarial attacks. Our findings confirm that PS effectively identifies flow-based NIDS features susceptible to problem-space perturbations. This novel approach leverages problem-space NIDS domain constraints as lightweight universal defense mechanisms against evasion adversarial attacks targeting flow-based ML-NIDS.
\end{abstract}


%PS thereby identifies features structurally resistant to evasion attacks in flow-based ML-NIDS due to the semantics of network traffic fields. This is because these features are constrained by domain-specific limitations and correlations in network traffic. 

% \fi
\IEEEpeerreviewmaketitle

\section{Introduction}
\label{Introduction}

%Machine Learning (ML) is widely employed in Network Intrusion Detection Systems (NIDS) due to its high accuracy in classifying large volumes of data \cite{el2023impact}. NIDS play a critical role in protecting computer networks by identifying malicious traffic. However,  ML-based NIDS models can be the target of evasion adversarial attacks \cite{ibitoye2019threat}. These attacks aim to deceive the ML model during decision-making by modifying or adding carefully crafted perturbations to the input data, often based on the gradient of the target ML model.

%\textbf{Contributions:} In this paper, we

%In the contemporary landscape of cybersecurity, Machine Learning (ML) has emerged as a cornerstone technology for Network Intrusion Detection Systems (NIDS). The ability of ML algorithms to process and classify vast amounts of network traffic data with high accuracy makes them indispensable for identifying malicious activities and safeguarding network integrity \cite{shehaby2023adversarial} \cite{ahmad2021network}. However, the effectiveness of ML-based systems is challenged by adversarial attacks, which exploit vulnerabilities in the ML models to fool their detection mechanisms \cite{ibitoye2019threat}. There are multiple types of adversarial attacks, such as poisoning \cite{cina2023wild}, backdoor \cite{saha2020hidden}, model stealing \cite{juuti2019prada}, and evasion attacks, which are the main focus of this paper. <-

%evasion adversarial attacks, which exploit vulnerabilities in the ML models to bypass detection mechanisms \cite{ibitoye2019threat}.

In the modern cybersecurity landscape, flow-based Machine Learning (ML) has emerged as a cornerstone technology for Network Intrusion Detection Systems (NIDS). The ability of ML algorithms to process and classify traffic at the flow level, where a flow represents aggregated packet metadata without inspecting payload content, with high speed and accuracy makes them indispensable for identifying malicious activities and safeguarding network integrity \cite{shehaby2023adversarial} \cite{ahmad2021network}. However, the effectiveness of ML-based systems is challenged by adversarial attacks, which exploit vulnerabilities in the ML models to fool their detection mechanisms \cite{ibitoye2025threat}. There are multiple types of adversarial attacks, such as poisoning \cite{cina2023wild}, backdoor \cite{saha2020hidden}, model stealing \cite{juuti2019prada}, and evasion attacks, the latter being the primary focus of this paper.
\begin{figure}[]
\centering
\includegraphics[width=1\linewidth,keepaspectratio=true]{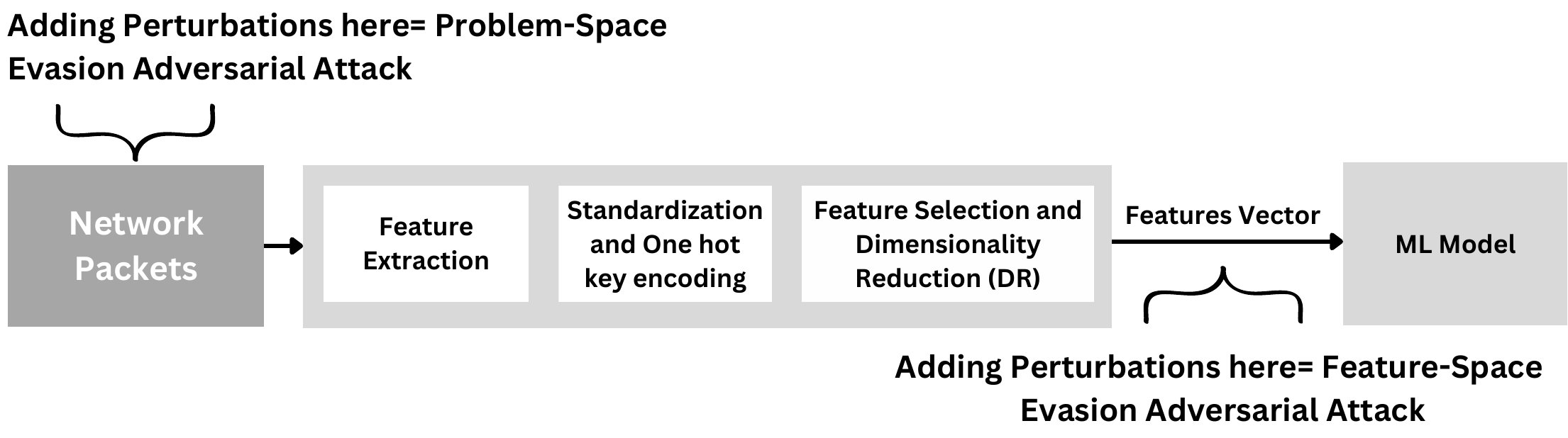}
	\caption{Evasion Adversarial Attacks in Feature-Space vs Problem-Space Against NIDS}
	\label{fig:ProvsFeat}
	\centering
\end{figure}
\begin{figure}[]
\centering
\includegraphics[width=1\linewidth,keepaspectratio=true]{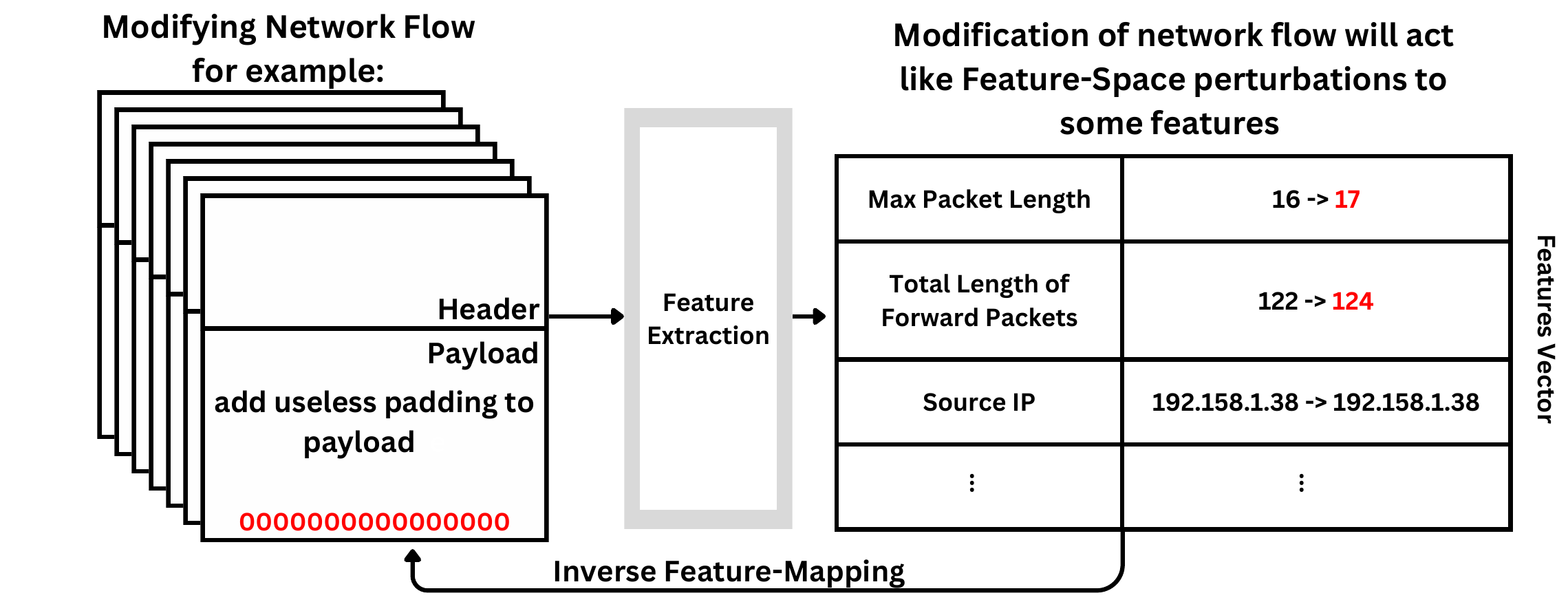}
	\caption{Example of Evasion Adversarial Attacks Problem-Space Perturbations Against NIDS}
	\label{fig:ProblemSpace}
	\centering
\end{figure}
%\subsection{Evasion adversarial attacks}
\textbf{Evasion adversarial attacks} involve the strategic manipulation of input data to deceive the ML model into making incorrect classifications. These perturbations are often crafted using gradients derived from the target model, allowing attackers to subtly alter the input while maintaining its functionality from a network perspective.

\subsection{ Feature-Space vs Problem-Space Evasion Adversarial Attacks Against ML-NIDS}

It is crucial to differentiate between feature-space and problem-space (real-world objects) \cite{biggio2013evasion} \cite{ibitoye2025threat}. %Feature-space refers to the representation of input data as features used by the ML model, while problem-space refers to the raw input data before feature extraction. Ibitoye et al. \cite{ibitoye2025threat} introduced the concept of 'space' in their taxonomy of adversarial attacks for network security, distinguishing between feature-space and problem-space attacks. 
\textbf{Feature-space adversarial attacks} manipulate or perturb feature vectors, assuming that the attacker can directly access and alter these features; however, this assumption is often unrealistic in practical scenarios. In contrast, \textbf{problem-space adversarial attacks} modify or perturb actual raw input data, such as network traffic flows (e.g., adding delays) or packets (e.g., padding the payload), which attackers can typically access. Fig.~\ref{fig:ProvsFeat} illustrates this distinction.

\begin{figure} 
\centering
\captionsetup[subfloat]{labelfont=scriptsize,textfont=scriptsize}
\centering
\subfloat[Two Examples of Perturb-able  Features\\ in Problem-Space Against NIDS]{%
\includegraphics[width=0.5\linewidth, keepaspectratio=true]{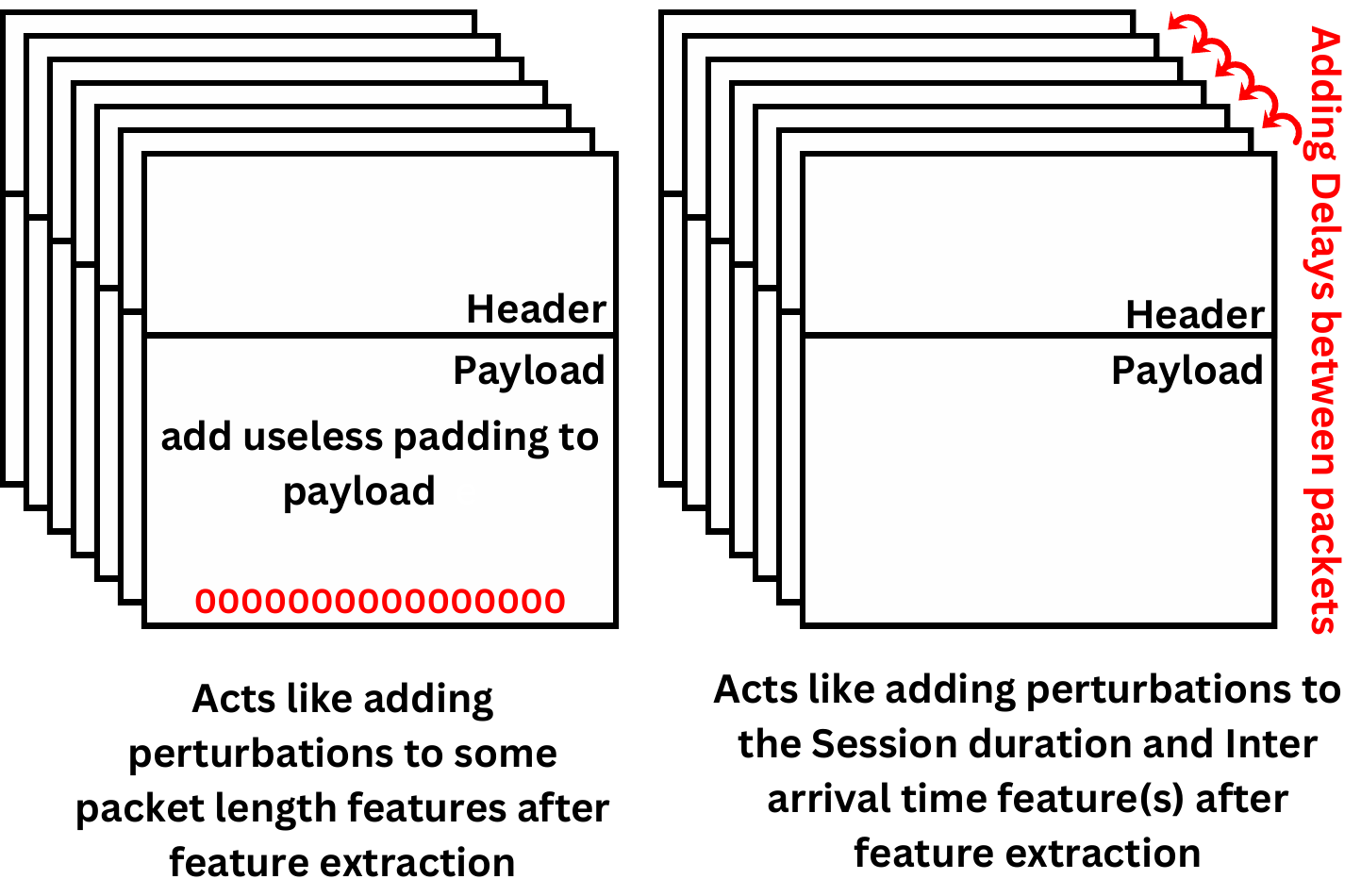}
\label{fig:Pert}
} 
\subfloat[Two Examples of Non-Perturb-able\\Features in Problem-Space Against NIDS]{%
\includegraphics[width=0.5\linewidth, keepaspectratio=true]{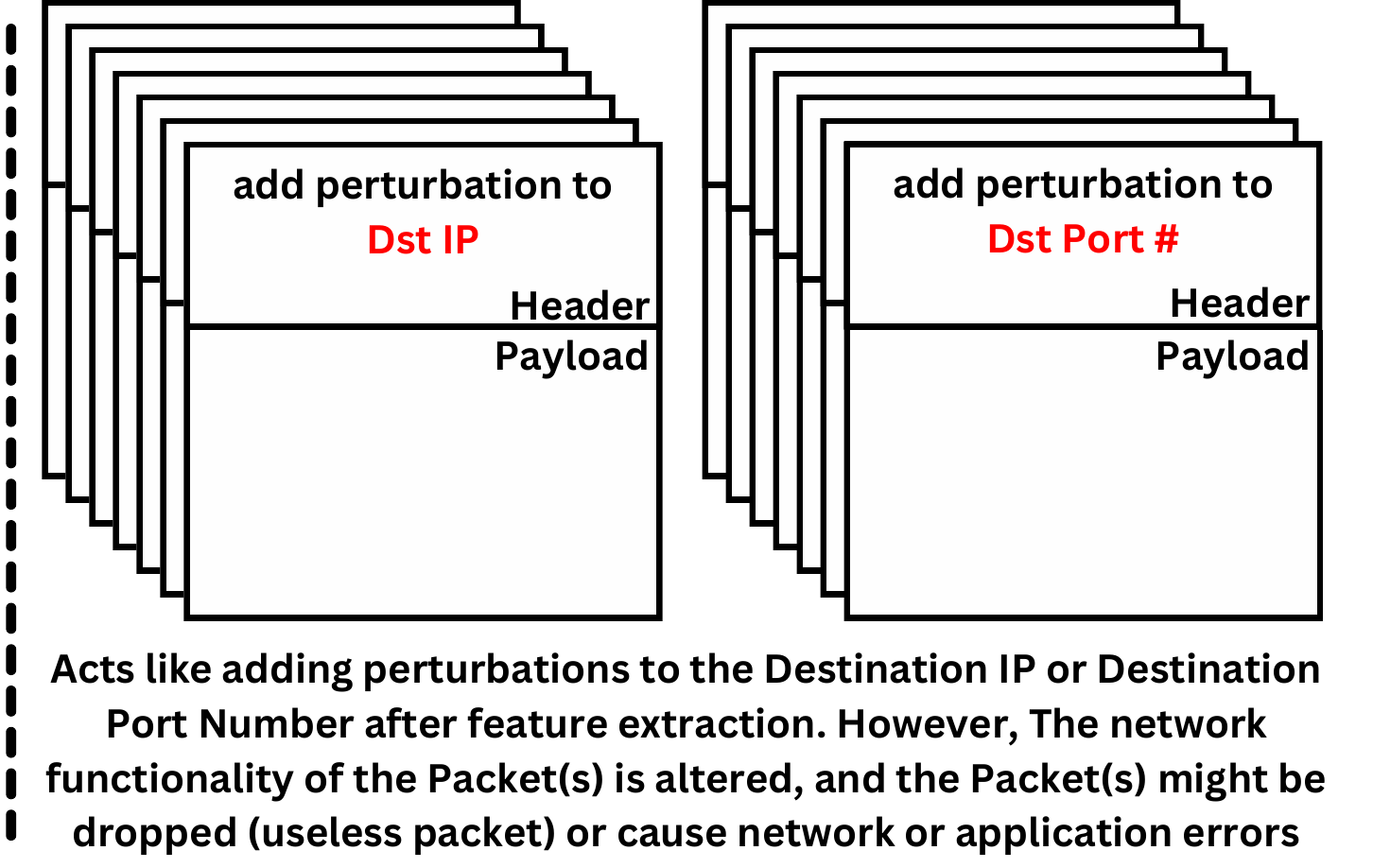}
\label{fig:nonPert}
}
\caption{Examples of Perturb-able vs Non-Perturb-able Features in Network Traffic}
\label{fig:PertNotPert}
\end{figure}

%In contrast, \textbf{problem-space adversarial attacks} modify or perturb actual raw input data, such as network traffic flow (e.g., adding delays) or packets (e.g., padding the payload), which attackers typically can access. Fig. ~\ref{fig:ProvsFeat} illustrates this distinction.
%modify actual data, such as network traffic (e.g., adding delays) or data packets (e.g., padding the payload), which attackers typically can access.
Feature-space attacks are often impractical against NIDS due to limited attacker access to feature vectors and complications from feature correlations and network constraints \cite{shehaby2023adversarial}. In contrast, problem-space attacks are more feasible, as external attackers can modify network packets directly. These attacks typically begin with feature-space perturbations and then translate them into real-world packet modifications (Inverse Feature-Mapping \cite{pierazzi2020intriguing}). As shown in Fig. \ref{fig:ProblemSpace}, attackers alter network flows to perturb targeted features after feature extraction, for example, adding payload padding to change the maximum packet length or total forward packet length. However, problem-space attacks also face challenges \cite{shehaby2023adversarial, grosse2024towards, apruzzese2023real}, including maintaining malicious objective and network functionality, keeping up with model and feature extraction updates, and adhering to NIDS feature constraints.
  %Feature-space attacks may not be practical against NIDS due to challenges an attacker would face in feature vector access and challenges with feature correlations and network constraints \cite{shehaby2023adversarial}. On the other hand, problem-space attacks are more practical than feature-space as the modifications happen to the network packets (feasible for an external attacker). They typically start with feature-space perturbations, then translate to real-world packet modifications (Inverse Feature-Mapping \cite{pierazzi2020intriguing}). In other words, as seen in Fig.~\ref{fig:ProblemSpace},  attackers modify the adversarial network flow to produce perturbations in certain targeted features within the feature vector after feature extraction. Fig.~\ref{fig:ProblemSpace} illustrates an example of evasion adversarial attack problem-space perturbations against NIDS, where the attacker adds padding to the payload of a packet to perturb features such as the maximum packet length and total length of forward packets in the feature vector. Despite being considered more practical than feature-space attacks, these attacks also face several practicality issues \cite{shehaby2023adversarial} \cite{grosse2024towards} \cite{apruzzese2023real}, such as: challenges in maintaining malicious objective and network functionality while altering packets; keeping up-to-date knowledge of the model, its features, and extraction techniques; or predicting correct common features. Problem-space attacks must also address NIDS features constraints.

\subsection{Perturb-ability of Features in Problem-Space Against NIDS}

Problem-space evasion attacks  on NIDS \cite{han2021evaluating, hashemi2019towards, vitorino2023sok, vitorino2023towards, vitorino2022adaptative, yan2023automatic, homoliak2018improving, apruzzese2024adversarial} involve modifying network packets to manipulate certain features within the feature vector. Perturbing some NIDS features in the problem-space without affecting network functionality might be feasible; for instance, adding padding to payloads or introducing delays between packets can perturb features such as length and interarrival time (Fig. \ref{fig:Pert}). However, problem-space constraints significantly limit the perturb-ability of many other NIDS features. For example, modifying the destination IP or port number disrupts the malicious capability or network functionality of the flow (Fig. \ref{fig:nonPert}), and certain features, like backward and inter-flow/connection features, are inaccessible for modification.

To address this distinction, \textbf{we coined the terms ``perturb-able'' and ``non-perturb-able'' features}. A perturb-able feature refers to a feature that can be altered through problem-space modifications without affecting the attacker's malicious capability or violating network constraints. Non-perturb-able, or robust, features, on the other hand, cannot be perturbed through such modifications without disrupting the malicious capability or network constraints. Fig. \ref{fig:PertNotPert} shows examples of perturb-able and non-perturb-able features in network traffic.

N.B. Some non-perturb-able (or robust) features may be completely unmodifiable due to problem-space limitations and correlations within the NIDS domain. For example, some backward features (features describing the network flow from server to client), like the mean size of a packet in the backward direction feature, can be extremely difficult for an attacker to access. However, it is important to note that most features can be modified through problem-space manipulations. By non-perturb-able features, we specifically refer to those that cannot be perturbed in the problem-space \textbf{while maintaining the attacker's malicious aim and complying with NIDS domain constraints.} For instance, changing the destination IP to manipulate its corresponding feature in the feature vector is possible. However, doing so would disrupt the flow's network functionality and the attacker's malicious objective, which is why we classify it as a non-perturb-able feature. 

\begin{figure}
%\footnotesize
\notsotiny
\centering
\begin{forest}
    for tree={
        grow=0,reversed, % tree direction
        parent anchor=east,child anchor=west, % edge anchors
      %  edge={line cap=round},outer sep=+1pt, % edge/node connection
       % rounded corners,minimum width=15mm,minimum height=8mm, % node shape
      %  l sep=10mm % level distance
    align=center
    }
  [\textbf{NIDS}\\\textbf{Features}
    [High\\Perturb-ability, L2 [Features can be perturbed in problem-space\\while maintaining NIDS constrains, L2]]
    [Medium\\Perturb-ability, L3[Features with medium possibility to be perturbed in \\problem-space while maintaining NIDS constrains,L3]]
    [Low\\Perturb-ability, L4[Features extremely hard to perturb in problem-space\\while maintaining NIDS constrains,L4]]
  ]
\end{forest}
  \caption{Classification of NIDS Features based on our proposed PS, where green represents a feature in the Low Perturb-ability class, yellow represents a feature in the Medium Perturb-ability class, and red represents a feature in the High Perturb-ability class}
	\label{fig:NIDSClass}

\end{figure}
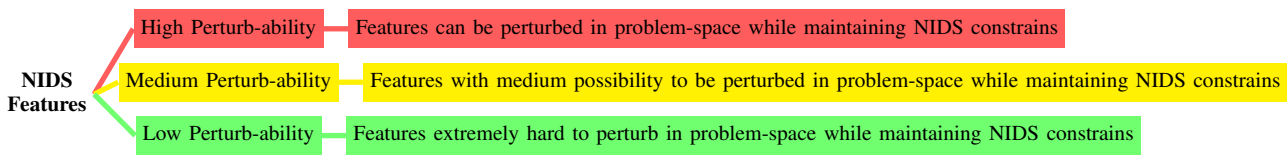

\subsection{Motivation and Aim}

Our motivation stems from the intuitive assumption that attackers can only access the problem-space rather than the feature-space. This perspective aligns with the reality of most network environments, where attackers can manipulate packet contents but do not have direct control over the feature extraction process (see \cref{ThreatModel} for more details on our threat model).

In response to this, our aim is to introduce the novel notion of the Perturb-ability Score (PS) metric, which is designed to enhance the robustness of ML-based NIDS. The PS metric helps to identify features in the problem-space that are susceptible to manipulation by attackers, without compromising the malicious or network functionality of traffic. By quantifying the perturb-ability of each feature within NIDS domain constraints, PS facilitates the selection of features that are inherently more resistant to adversarial attacks or the masking of nonresistant features. Our aimed classification is shown in Fig. \ref{fig:NIDSClass}. For the remainder of the paper, we will use the color scheme found in Fig. \ref{fig:NIDSClass}, i.e., a green feature represents a feature in the Low Perturb-ability class, a yellow feature represents a feature in the Medium Perturb-ability class, and a red feature represents a feature in the High Perturb-ability class.

%Our approach offers highly effective defenses against evasion adversarial attacks targeting ML-based Network Intrusion Detection Systems (ML-NIDS).

What sets our proposed defenses apart is their independence from attack types, attack norms \cite{carlini2017towards} used, or the level of adversarial knowledge (whether black-box, white-box, or gray-box \cite{ibitoye2025threat}). Unlike most conventional defenses that focus primarily on the internal mechanisms of the ML model itself, our method takes an \textit{\textbf{"outside-the-box"}} perspective. By leveraging inherent network domain constraints external to the ML model, we significantly reduce the attack surface. This strategic shift from model-centric defenses to exploiting network domain properties introduces a novel layer of protection, enhancing the overall robustness of ML-NIDS against adversarial threats.

\subsection{Contributions}

Our contributions are threefold: \textbf{\circled{1} Introduction of Perturb-ability Score (PS) (\cref{PS}):} We propose a novel Perturb-ability Score (PS) metric to quantify the vulnerability of NIDS features to adversarial manipulation. The PS measures the susceptibility of a feature to perturbation by an attacker without compromising the underlying malicious objective or network functionality of the attack. This metric establishes a robust foundation for assessing feature resilience in ML-NIDS adversarial contexts. \textbf{\circled{2} Leveraging PS in Defensive Mechanisms (\cref{def}):} By utilizing PS, we introduce defense strategies that enhance the robustness of ML-based NIDS. \textbf{\circled{2a} PS-guided feature selection (\cref{defA})} enables the selection of only inherently resilient features that remain robust against adversarial perturbations during pre-processing. This approach fortifies NIDS by reducing the attack surface while maintaining solid detection performance across various models and datasets. \textbf{\circled{2b} PS-guided feature masking (\cref{defB})} of high perturb-ability features, which also narrows the attack surface by replacing easily perturb-able features with neutral values, effectively removing the attacker's ability to manipulate those features while maintaining the model's structure and dimensionality. \textbf{\circled{3} Mapping Problem-Space Adversarial Attacks to Feature-Space (\cref{map}) and Conducting Thorough Testing (\cref{class,perf,advtest}):} We conduct an in-depth analysis of problem-space adversarial attack techniques found in the literature by mapping their traffic morphing methods to the corresponding NIDS features. This mapping validates the PS classification, demonstrating how PS effectively captures the impact of problem-space evasion techniques on NIDS features and provides significant insights into how adversarial attacks manifest in both problem and feature-spaces. We also conducted meticulous testing and comparison using feature-space and problem-space attacks, comparisons using three public datasets, multiple models and comparisons with adversarial training.

\section{Threat Model}
\label{ThreatModel}
The following is our threat model; \textbf{Target:} We assume that the target model is a flow-based NIDS, rather than packet-based or raw traffic-based NIDS.  
\textbf{Attacker's Knowledge:} We assume that the attacker has no knowledge of the selected features (feature vector) used by the model. However, other information, such as the ML algorithm and activation functions, may be known to the attacker.
\textbf{Attacker's Capability:} We assume that the attacker does not have access to the feature vector but can only access and alter network packets or flows (problem-space modifications). In the NIDS context, perturbing network flow at a packet granular level may be challenging if it results in a network flow that does not correspond to any valid network behavior. Thus, we assume that attackers might deploy innovative approaches to crafting problem-space attacks by changing flow behavior or manipulating hosts' behavior instead of directly modifying packets, for example, adjusting the output rate, adding packet delay, introducing packet loss, and implementing duplication \cite{catillo2024towards}. Additionally, attackers must be capable of creating adversarial flows while adhering to several constraints: (1) Maintaining Functionality: The malicious capability must be preserved alongside the network functionality of the packets. (2) NIDS Feature Constraints: The attacker needs to consider the feature limitations of the NIDS model. These constraints might involve specific data formats, packet sizes, correlations, or limitations on certain network protocols. \textbf{Attacker's Goal:} The attacker aims to compromise the integrity of the NIDS by evading detection (evasion adversarial attacks) while maintaining the malicious functionality and operational integrity of the perturbed network flow.
%\textbf{Attacker's Capability:} We assume that the attacker does not have access to the feature vector. Instead, the attacker can only access and alter network packets or flows (problem-space attack).  

Although recent research has cast doubt on the practicality of evasion adversarial attacks against NIDS \cite{shehaby2023adversarial, apruzzese2022modeling, el2023impact}, since knowing the ML algorithm, activation functions, or querying the NIDS is unlikely, this paper nonetheless focuses on exploring the possibility and implications of adding perturbations to NIDS features in the problem-space.

%Despite the critical importance of defending against adversarial attacks, much of the existing research has focused on computer vision datasets, where the primary objective is to fool human perception \cite{goodfellow2014explaining}. This focus has led to the development of algorithms optimized for visual data, which may not directly translate to the network security domain. Furthermore, many studies on adversarial attacks against ML-based NIDS operate under the assumption that adversaries have full control over the feature-space, an unrealistic scenario in practical settings \cite{biggio2013evasion}. 

\section{Perturb-ability Score (PS)}
\label{PS}

In this section, we explain how our novel Perturbability Score (PS) quantifies the susceptibility of each flow-based NIDS feature to problem-space evasion attack perturbation.

\subsection{NIDS Features Perturb-ability Classification}
The aim of our perturb-ability Score (PS) is to classify NIDS features based on their susceptibility to perturbations within the problem-space while adhering to NIDS constraints. By NIDS constraints, we refer to the problem-space constraints within the NIDS domain, including the limitations, correlations, and restrictions inherent to network traffic and NIDS attacks. This classification is crucial for understanding the robustness of NIDS against evasion adversarial attacks and may be utilized as a defense, as we will discuss later in this paper. As seen in Fig. \ref{fig:NIDSClass}, our PS aims to categorize features into three main groups: high Perturb-ability, medium Perturb-ability, and low Perturb-ability.

High Perturb-ability features (high PS) can be perturbed in the problem-space while adhering to NIDS domain constraints, such as the maximum inter-arrival time (IAT) between packets in the forward direction (from client to server). In contrast, low Perturb-ability features are difficult or extremely difficult to perturb while maintaining these constraints, like the destination IP. Medium Perturb-ability features fall between these two extremes.

\subsection{PS Formulation}

%Evaluation Criteria of PS}

The goal of PS is to obtain a Perturb-ability score for each feature ($f_i$) in a dataset $D$, where $i$ is the ID of the feature from 1 to $n$, and $n$ is the number of features in $D$. PS should range from 0 (features extremely hard to perturb in problem-space while maintaining the networks constraints) to 1 (features can be perturbed in problem-space while maintaining the networks constraints). The $PS_{Total}[f_i]$ is the geometric average of the following five fields:
\begin{comment}

\begin{figure}
\centering
\includegraphics[width=0.7\linewidth,keepaspectratio=true]{Pics/PS3_plot_5.png}
	\caption{Visualization of the PS\textsubscript{3}[$f_i$] Evaluation Equation}
	\label{fig:PS3}
	\centering
\end{figure}

\end{comment}

\begin{figure}
\centering
\captionsetup[subfloat]{labelfont=scriptsize,textfont=scriptsize}
\subfloat[{\scriptsize Visualization of the PS\textsubscript{2}[$f_i$] Evaluation Equation, where MaxR=255 and MinR=2}]{%
    \includegraphics[scale=1, width=0.45\linewidth, height=0.22\linewidth]{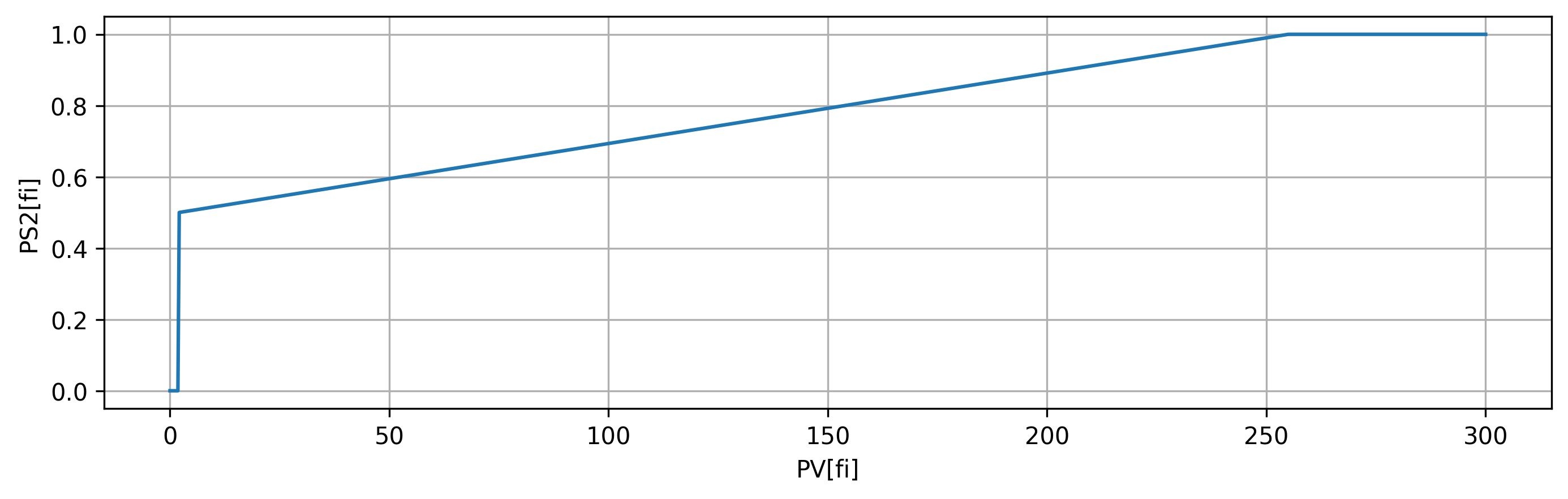}
    \label{fig:PS2}
}%
\hfill
\subfloat[{\scriptsize Visualization of the PS\textsubscript{3}[$f_i$] Evaluation Equation}]{%
    \includegraphics[scale=1, width=0.45\linewidth, height=0.22\linewidth]{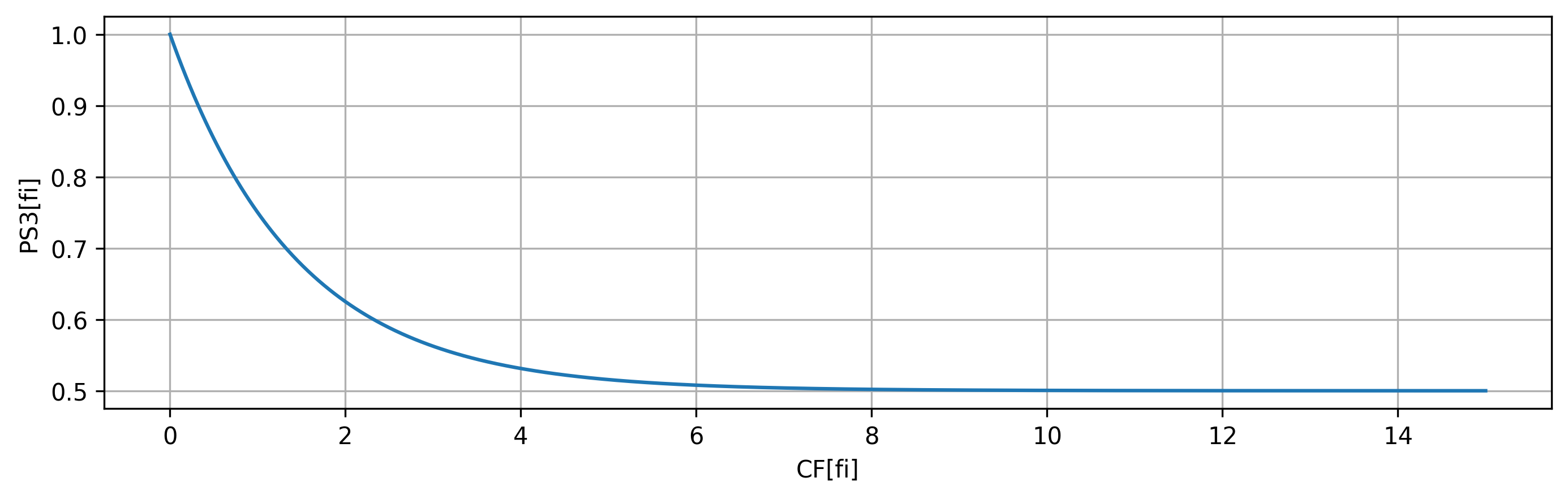}
    \label{fig:PS3}
}

%{Pics/PS3_plot_5.png}
\caption{Visualizations of the PS\textsubscript{2}[$f_i$] and PS\textsubscript{3}[$f_i$] Evaluation Equations}
\label{fig:PS_combined}
\end{figure}

\subsubsection{PS\textsubscript{1}[$f_i$]: Critical Protocol, Identifiers and Functional Integrity Felids}

%strict Header features and network/malicious functionality of network flows after adding perturbations in the problem-space.

%(IP addresses in TCP flows, destination port number or protocol)

This PS field focuses on critical features where perturbations in the problem-space could impact network functionality or enable malicious behavior in network flows.

PS\textsubscript{1}[$f_i$] will be 0 if any of the following conditions are true (which will make PS\textsubscript{Total}[$f_i$] equals 0);

\textbf{C1:} the feature $f_i$ is protocol identification feature. 

\textbf{C2:} the feature $f_i$ is a critical address/identifier (e.g., IP address in a TCP flow, port number).

\textbf{C3:} the feature $f_i$ is a field that defines a network functional integrity attribute (e.g., Service, State, flow direction)

%address/identifier (IP address, port number, etc.).

%adding perturbation to feature $f_i$ will affect the network or malicious functionality of the flow.

PS\textsubscript{1}[$f_i$] can be described with the following equation:
\[
    PS\textsubscript{1}[f_i]= 
\begin{cases}
    0,& \text{if } \text{(C1 or C2 or C3)}\\
    1,              & \text{otherwise}
\end{cases}
\]

%\subsubsection{PS\textsubscript{2}[$f_i$]: The range of Possible Values of a Feature:\newline}

\subsubsection{PS\textsubscript{2}[$f_i$]: The range of Possible Values of a Feature} This PS field considers the cardinality (number of possible values) of a NIDS feature. In unconstrained domains like computer vision, attackers can freely perturb pixels, which typically have a range of 0 to 255 per channel (e.g., red, green, blue). Conversely, certain NIDS features have limited cardinality. For example, a NIDS dataset may have binary or categorical features with a limited number of categories. Such features offer less flexibility to attackers. The gradients of the targeted model might suggest perturbations in a specific direction, but the attacker might be unable to comply due to the limited number of possible feature values of these features.

%feature where its value is 1 if the source and destination IP addresses are equal and the source and destination port numbers are equal; otherwise, its value is 0.

%PS\textsubscript{2}[$f_i$] will be 1 if $f_i$'s number of Possible Values ($PV$) is greater than 255 (this feature will be similar to computer vision's pixel, and it will be flexible to perturb). On the other hand, if $f_i$'s $PV$ ($PV$[$f_i$]) is less than or equal to 255, PS\textsubscript{2}[$f_i$] will be equal to a linear function where its output is 1 if $f_i$'s $PV$ is 255, and 0.5 if $f_i$'s $PV$ is 2 (binary). If $PV[f_i]$ is less than 2 (equals 1), it indicates that $f_i$ is non-perturb-able, in which case $\text{PS}_2[f_i]$ will be set to 0. However, in this case (where $PV[f_i]$ equals 1), we recommend dropping that feature, as it does not contribute meaningful information to the ML model. Fig. \ref{fig:PS2} shows the visualization of the PS\textsubscript{2}[$f_i$] evaluation equation. 

PS\textsubscript{2}[$f_i$] will be 1 if $f_i$'s number of Possible Values ($PV$) is greater than $\text{MaxR}$, where we set $\text{MaxR} = 255$ (this threshold is chosen because features with this many possible values behave similarly to computer vision's pixels, offering significant flexibility for perturbation). On the other hand, if $f_i$'s $PV$ ($PV$[$f_i$]) is less than or equal to $\text{MaxR}$, PS\textsubscript{2}[$f_i$] will be equal to a linear function where its output is 1 if $f_i$'s $PV$ is $\text{MaxR}$, and 0.5 if $f_i$'s $PV$ is $\text{MinR}$, where we set $\text{MinR} = 2$ (representing binary features). If $PV[f_i]$ is less than $\text{MinR}$ (equals 1), it indicates that $f_i$ is non-perturb-able, in which case $\text{PS}_2[f_i]$ will be set to 0. However, in this case (where $PV[f_i]$ equals 1), we recommend dropping that feature, as it does not contribute meaningful information to the ML model. Fig. \ref{fig:PS2} shows the visualization of the PS\textsubscript{2}[$f_i$] evaluation equation. It is worth noting that practitioners and domain experts can calibrate the $\text{MinR}$ and $\text{MaxR}$ thresholds based on specific application requirements and the nature of the feature-space under consideration.

%If $PV$[$f_i$] is less then 2, that means that $f_i$ is non-perturb-able, so in that case, PS\textsubscript{2}[$f_i$] will be 0. However, If $PV$[$f_i$] is less then 2 (equals 1), we recommend of droppinf that features as it does not represent any information to the ML model. 

PS\textsubscript{2}[$f_i$] can be described with the following equation:

\[
\text{PS}_2[f_i] = \begin{cases} 
1 & \text{if } \text{PV[$f_i$]} > \text{MaxR} \\
0 & \text{if } \text{PV[$f_i$]} < \text{MinR} \\
0.5 + (0.5 \times \frac{(\text{PV[$f_i$]} - \text{MinR})}{(\text{MaxR} - \text{MinR})}) & \text{otherwise} 
\end{cases}
\]

%{\color{red} Small perturbations are not crucial here.
%Standardization/Normalization}

Some might argue that the perturbations in evasion adversarial attacks are minuscule \cite{goodfellow2014explaining}, and therefore, the cardinality of the feature should not affect its perturb-ability. While this may hold true in domains like computer vision, where adversarial perturbations are optimized for human perception and must remain imperceptibly small, it is not the case for network data \cite{sheatsley2022adversarial}. In network security, small perturbations often have limited to no utility. In other words, the similarity constraint \cite{he2023adversarial}, which ensures that adversarial examples are nearly indistinguishable from the original examples in domains like computer vision, is not applied in the feature-space of attacks against ML-NIDS. Instead, the similarity constraint is placed on the semantics of the attack. Consequently, adversarial attacks in the problem-space can introduce significantly larger perturbations to the features \cite{he2023adversarial}. Moreover, since features are typically normalized or standardized, altering the value of a feature in the problem-space may require even larger perturbations. Thus, the cardinality of a feature becomes a critical factor for an attacker attempting to craft successful evasion adversarial attacks against ML-NIDS in the problem-space.

\subsubsection{PS\textsubscript{3}[$f_i$]: Correlated Features}

This PS field considers the correlation between a NIDS feature and other features. Due to network constraints within NIDS, many features exhibit problem-space correlations. For instance, the flow duration feature is typically correlated with the total forward and backward inter-arrival times. Such correlated features limit the attacker's flexibility. The gradients of the targeted model might recommend a specific perturbation to one feature and a different perturbation to another. However, achieving these opposing perturbations simultaneously is very difficult if the features are highly correlated within the problem-space. As an example, an attacker cannot simultaneously increase the flow duration while decreasing both the forward and backward inter-arrival times. As the number of correlated features associated with a single feature increases, it becomes more difficult to perturb that feature in the problem-space. These correlations also cause collateral damage effects, which we analyze in Section \ref{map}.

%As the number of correlated features increases per feature, it becomes harder to perturb that feature in the problem-space.

%When two features are correlated, an attacker attempting to perform adversarial attacks faces a specific challenge: the probability that gradients will mandate movement in the same direction is 50 percent. This probability decreases exponentially as the number of correlated features increases. 

%When two features are correlated, there is a 50\% probability that gradients will mandate perturbations in the same direction (either both gradients align or oppose each other). Each additional correlated feature doubles the number of possible gradient configurations and halves the probability of the gradients requiring perturbations in similar direction. As a result, this probability decreases exponentially as more correlated features are added. To account for this phenomenon in our PS\textsubscript{3}[$f_i$] calculation, we implement an exponential formula that appropriately weights the correlation effects.  Fig. \ref{fig:PS3} shows the visualization of the PS\textsubscript{3}[$f_i$] evaluation equation.

When two features are correlated, there is a 50\% probability that gradients will require perturbations in the same direction (either both gradients align with or oppose each other). Each additional correlated feature doubles the number of possible gradient configurations and halves the probability of the gradients requiring perturbations in a similar direction. As a result, this probability decreases exponentially as more correlated features are added. To account for this phenomenon in our PS\textsubscript{3}[$f_i$] calculation, we introduce an exponential formula that appropriately weights the correlation effects. Fig.~\ref{fig:PS3} illustrates the PS\textsubscript{3}[$f_i$] evaluation equation.

%\textcolor{red}{PS\textsubscript{3}[$f_i$] will follow a linear function, where its output is 0.5 if the number of Correlated Features ($CF$) of $f_i$ is equal to or greater than a threshold (the maximum number observed in our experiments was 10, which we chose as the threshold), and 1 if $f_i$'s $CF$ (CF[$f_i$]) is 0.  Fig. \ref{fig:PS3} shows the visualization of the PS\textsubscript{3}[$f_i$] evaluation equation.}

PS\textsubscript{3}[$f_i$] can be described with the following equation, where CF[$f_i$] represents the number of correlated features to feature $f_i$:

\begin{comment}

\textcolor{red}{
\[
\text{PS}_3[f_i] = 1 - 0.05 \times \min(\text{CF[$f_i$]}, 10)
\]
}
\end{comment}
\[
\text{PS}_3[f_i] = 0.5 + (0.5 \times \frac{1}{2^\text{CF[$f_i$]}})
\]

We have deliberately constrained the range of values from 1 to 0.5 to ensure that even in scenarios with numerous correlated features, the PS\textsubscript{Total} value is only slightly reduced. This design choice acknowledges that some attacks are blind in nature and do not rely on gradient information, making them less susceptible to correlation effects. The exponential formula therefore provides a balanced approach that maintains defensive integrity while accurately modeling the diminishing impact of correlation on attack success probability.

\begin{comment}

\[
\text{PS}_3[f_i] =
\begin{cases} 
1 & \text{if } \text{CF[$f_i$]} < 2 \\
1 - 0.05 \times \min(\text{CF[$f_i$]}, 10) & \text{if } \text{CF[$f_i$]} \geq 2
\end{cases}
\]

\end{comment}
%=1-0.05*MIN(K2,10)

\textbf{Handling NIDS feature correlations:} The ADAPTIVE-JSMA (AJSMA) attack \cite{sheatsley2022adversarial} is an enhanced version of the Jacobian-based Saliency Map Approach (JSMA), specifically adapted for network intrusion detection systems with domain-specific constraints. This attack allows perturbations to be applied in either direction (increasing or decreasing the feature's value), depending on which direction will push the input closer to the target class while adhering to the domain constraints. However, AJSMA focuses on techniques for attacking NIDS at the feature layer, rather than in the problem-space. The complexity of real-world network environments is significantly higher than constraints modeled in the feature layer, such as protocol-related constraints (TCP, UDP, etc.), which are relatively simple compared to the extensive limitations present in the real-world problem-space like side effect features \cite{pierazzi2020intriguing}.

%, which are not cons (e.g., injecting new packets into the traffic or fragmenting packets can cause new features to be extracted),}

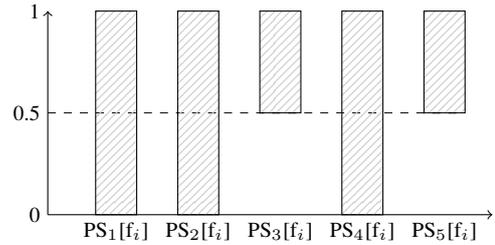
\begin{figure}
\footnotesize
\centering
\begin{tikzpicture}[scale=0.8]
    % Define custom pattern
    \tikzset{
        diagonal lines/.style={pattern=north east lines, pattern color=gray!40}
    }
    % Draw axes
    \draw[->] (0,0) -- (6.5,0) node[right] {};
    \draw[->] (0,0) -- (0,3) node[above] {};
    
    % Y-axis labels
    \node[left] at (0,0) {0};
    \node[left] at (0,1.5) {0.5};
    \node[left] at (0,3) {1};
    
    % Draw dashed line at 0.5
    \draw[dashed] (0,1.5) -- (6,1.5);
    
    % Draw narrower bars with diagonal lines
    \fill[diagonal lines] (0.7,0) rectangle (1.3,3);
    \fill[diagonal lines] (1.9,0) rectangle (2.5,3);
    \fill[diagonal lines] (3.1,1.5) rectangle (3.7,3);
    \fill[diagonal lines] (4.3,0) rectangle (4.9,3);
    \fill[diagonal lines] (5.5,1.5) rectangle (6.1,3);
    
    % Draw narrower bar outlines
    \draw (0.7,0) rectangle (1.3,3);
    \draw (1.9,0) rectangle (2.5,3);
    \draw (3.1,1.5) rectangle (3.7,3);
    \draw (4.3,0) rectangle (4.9,3);
    \draw (5.5,1.5) rectangle (6.1,3);
    
    % X-axis labels with subscript i
    \node[below] at (1,0) {PS$_1$[f$_i$]};
    \node[below] at (2.2,0) {PS$_2$[f$_i$]};
    \node[below] at (3.4,0) {PS$_3$[f$_i$]};
    \node[below] at (4.6,0) {PS$_4$[f$_i$]};
    \node[below] at (5.8,0) {PS$_5$[f$_i$]};
\end{tikzpicture}
\caption{Ranges of PS Fields}
	\label{fig:Ranges}
	\centering
\end{figure}

Nevertheless, we acknowledge that an attacker with exceptional knowledge of the attacked ML model and its feature vector could potentially overcome the constraints posed by correlated features in the problem-space. This is another reason why the minimal value of PS\textsubscript{3}[$f_i$] is set to 0.5, resulting in only a modest reduction in PS\textsubscript{total}[$f_i$]. However, we still believe that correlated features in the problem-space would pose a significant challenge for an attacker in a practical setting where they have no information about the attacked ML-NIDS and cannot query it. Additionally, as the number of correlated features increases, it becomes exponentially harder for an attacker to handle these correlations.

\subsubsection{PS\textsubscript{4}[$f_i$]: Features that attackers cannot access}
This PS field focuses on features that attackers cannot access. Examples of such features include backward features (e.g., Minimum Backward Packet Length) and interflow features (e.g., number of flows that have a command in an FTP session (ct\_ftp\_cmd)).

PS\textsubscript{4}[$f_i$]'s value will depend on the following conditions;

\textbf{C4:} the feature $f_i$ is not a backward or interflow feature. In other words, attackers can access $f_i$.

\textbf{C5:} the feature $f_i$ is a backward or interflow feature; however, it is highly correlated with a forward feature. In other words, attackers can modify $f_i$ in an indirect way. 

\textbf{C6:} the feature $f_i$ is a backward or interflow feature; however, it is correlated with multiple forward features. In other words, attackers can modify $f_i$ indirectly, but it will be challenging for them as it is correlated with multiple features.

\textbf{Otherwise (if none of C4, C5, or C6 apply):} the feature $f_i$ is a backward or interflow feature and it is not correlated with any forward feature. In other words, attackers cannot access $f_i$.

\[
    PS\textsubscript{4}[f_i]= 
\begin{cases}
    1,& \text{if } \text{(C4 or C5)}\\
    0.5,& \text{if } \text{(C6)}\\
    0, & \text{otherwise}
\end{cases}
\]

\subsubsection{PS\textsubscript{5}[$f_i$]: Flow-wide Correlation}

%This PS field considers features that are correlated with numerous flow packets.

This PS field considers whether modifying a feature requires altering the entire flow. Features like mean or standard deviation, which depend on multiple packets within a flow, are harder to perturb due to their broader impact on the network flow.

PS\textsubscript{5}[$f_i$]'s value will depend on the following condition;

\textbf{C7:} $f_i$ is a feature that requires modifying the entire flow of packets (forward, backward, or both), such as mean or standard deviation features.

\[
    PS\textsubscript{5}[f_i]= 
\begin{cases}
    0.5,& \text{if } \text{(C7)}\\
    1, & \text{otherwise}
\end{cases}
\]

\subsubsection{PS\textsubscript{Total}[$f_i$]}
The overall Perturb-ability Score (PS\textsubscript{Total}[$f_i$]) for each feature $f_i$ is calculated as the geometric mean of the five individual PS fields we defined. These PS fields are assigned a value of 0 if a specific condition renders feature $f_i$ non-perturb-able within the problem-space. A value of 0.5 is assigned if a condition only reduces the feasibility of perturbing $f_i$. The geometric mean was chosen to ensure that PS\textsubscript{Total}[$f_i$] becomes 0 if any of the individual PS fields have a value of 0. %This reflects the fact that a single highly limiting PS field can significantly reduce the overall perturb-ability of a feature. 
However, it's important to note that any PS field value below 1 will contribute to a decrease in the overall PS\textsubscript{Total}[$f_i$].

PS\textsubscript{Total}[$f_i$] can be described with the following equation:

\[
\text{PS\textsubscript{Total}[$f_i$]} = \sqrt[5]{\prod_{j=1}^{5} \text{PS\textsubscript{j}[$f_i$]}}
\]

The PS$_\text{Total}$ will be calculated for all features $f_i$ in the dataset, from $i = 1$ to $n$, where n is the number of features in the dataset.

\begin{comment}

\begin{figure}
\centering
\includegraphics[width=1\linewidth,keepaspectratio=true]{samples/Pics/CORRM15.png}
	\caption{The correlation Matrix between the features in UNSW-NB15 Dataset}
	\label{fig:CORRMCSE}
	\centering
\end{figure}

\begin{figure}
\centering
\includegraphics[width=1\linewidth,keepaspectratio=true]{samples/Pics/CORRMCSE.png}
	\caption{The correlation Matrix between the features in CSE-CIC-IDS2018 Dataset}
	\label{fig:corrmcse}
	\centering
\end{figure}

\begin{figure}
\centering
\includegraphics[width=1\linewidth,keepaspectratio=true]{samples/Pics/CorrUNSW.png}
	\caption{The correlated features in UNSW-NB15 Dataset}
	\label{fig:CorrUNSW}
	\centering
\end{figure}

\begin{figure}
\centering
\includegraphics[width=1\linewidth,keepaspectratio=true]{samples/Pics/corrCSE.png}
	\caption{The correlated features in CSE-CIC-IDS2018 Dataset}
	\label{fig:corrCSE}
	\centering
\end{figure}

\end{comment}

\subsection{Ranges of PS Fields and Thresholds Calibration}

Fig.~\ref{fig:Ranges} illustrates the possible ranges of the five PS fields. The ranges for $\text{PS}_1[f_i]$, $\text{PS}_2[f_i]$, and $\text{PS}_4[f_i]$ span from 0 to 1, while $\text{PS}_3[f_i]$ and $\text{PS}_5[f_i]$ exhibit more restricted ranges, from 0.5 to 1. This variation in the range of PS values reflects the nuanced characteristics of each PS field's vulnerability to adversarial perturbation.

%As seen in Figure \ref{fig:Ranges}, visualizes the possible ranges of the 5 PS fields, the ranges for $\text{PS}_1[f_i]$, $\text{PS}_2[f_i]$, and $\text{PS}_4[f_i]$ span from 0 to 1, while $\text{PS}_3[f_i]$ and $\text{PS}_5[f_i]$ exhibit more restricted ranges, from 0.5 to 1. This variation in the range of PS values reflects the nuanced characteristics of each PS field's vulnerability to adversarial perturbation.

As previously discussed, the total perturb-ability score (PS\textsubscript{Total}[$f_i$]) for each feature $f_i$ is computed as the geometric mean of the five individual PS fields, ensuring a balanced assessment across different conditions. This approach was adopted to account for the fact that certain feature properties can entirely negate their perturb-ability, such as when the feature is inaccessible to the attacker. These features are given a PS of 0. On the other hand, other conditions may only reduce the feature’s PS without fully eliminating its susceptibility. For instance, features that require altering an entire packet flow, whether forward, backward, or both, such as mean or standard deviation of packet properties, tend to decrease the overall PS score without completely nullifying it.
\begin{figure}
\centering
\includegraphics[width=0.9\linewidth,keepaspectratio=true]{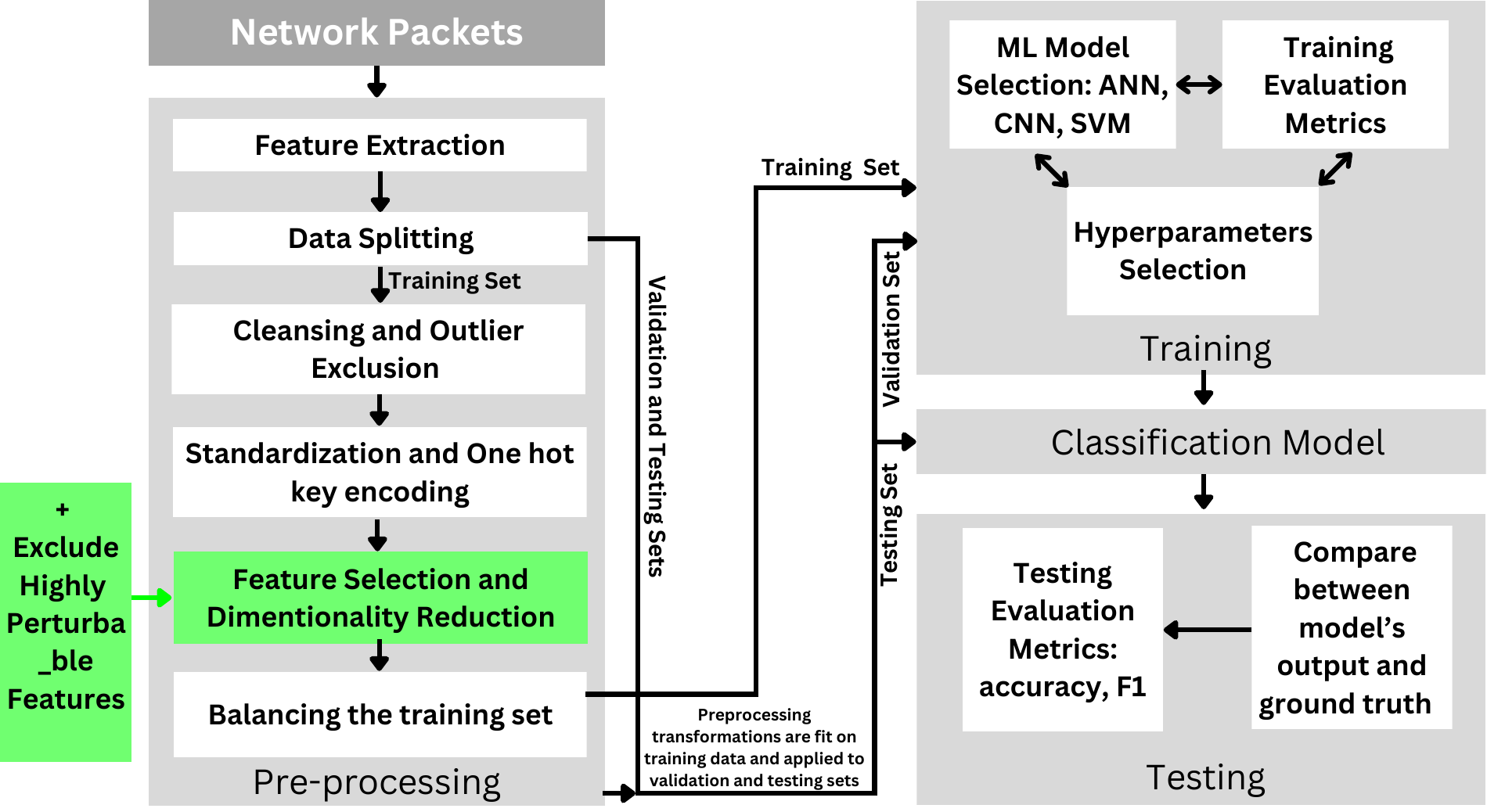}
	\caption{Option A Defense: PS-enabled Feature Selection}
	\label{fig:PSAsaDefense}
	\centering
\end{figure}

%Using PS as a Potential Defense against Practical Problem-Space Evasion Adversarial Attacks
The decision to define varying ranges for different PS fields introduces an implicit weighting mechanism, where each field contributes to the final PS score with varying significance. This differential treatment mirrors real-world scenarios, where certain feature characteristics inherently have a greater influence on the overall perturb-ability than others. Hence, the structured variation in PS field ranges allows for a more precise and context-sensitive evaluation of each feature’s robustness.

While certain thresholds, such as 2 ($\text{MinR}$) and 255 ($\text{MaxR}$) in $\text{PS}_2[f_i]$, might appear arbitrary at first glance, in the previous section, we provided a detailed rationale for these choices to the best of our ability. However, it is important to recognize that these thresholds, along with the underlying functions of the PS fields, are not set in stone. As with many design threshold decisions, these parameters and thresholds can, and often should, be calibrated, adapted and fine-tuned by machine learning engineers, domain experts, and practitioners prior to deployment. The flexibility to adjust such thresholds ensures that the perturb-ability scoring system remains both robust and adaptable to varying and dynamic real-world scenarios, where the specific characteristics of the network and potential threat models may differ.

\section{PS-enabled Defenses}
\label{def}

In this section, we introduce two methods where PS enables defenses against practical problem-space evasion adversarial attacks against ML-NIDS.

\subsection{Option A: PS-enabled Feature Selection}
\label{defA}
%during The development of the flow-based ML-NID
Leveraging feature constraints in Network Intrusion Detection Systems offers a promising defense against problem-space adversarial attacks. Fig.~\ref{fig:PSAsaDefense} presents our novel defense mechanism, which integrates the Perturb-ability Score (PS) as a key component of the feature selection process.

By the exclusion of features with high perturb-ability scores during the feature selection process during the development of an ML-NIDS, attackers encounter no or very few perturb-able features in problem-space, significantly reducing the attack surface and making it significantly more difficult for adversaries to exploit the system. This method ensures that the features retained for training and classification are inherently resistant to adversarial manipulations. While this may require rethinking traditional feature selection methods, the potential benefits in preventing evasion attempts are substantial. This simple, efficient solution utilizes NIDS domain constraints as a defense with minimal computational overhead.

\subsection{Option B: PS-enabled Feature Masking}
\label{defB}

%\section{Option B: Feature Masking for High PS Features}

%As illustrated in Figure \ref{fig:PSAsaDefenseOptionB}, 

Unlike our Option A defense, which requires retraining the entire model on a selected subset of features if the model is already running, Option B provides a more efficient approach when the system is already deployed and the cost of re-selecting features and re-training is prohibitively high. Our novel approach allows us to maintain the existing model architecture and dimensionality while masking high perturb-ability features.

Our feature masking approach builds upon the binary feature mask optimization framework introduced by Lorasdagi et al. \cite{lorasdagi2024binary}, but with a critical distinction. Their approach is primarily for feature selection during model development, while ours is designed as a defense mechanism for already deployed systems.

We define a mask vector $\mathbf{m} \in {0,1}^M$, where $M$ is the number of features in the dataset. The mask is determined by the PS scores:

\[ 
m_i =
\begin{cases}
0, & \text{if } \text{PS}_{\text{Total}}[f_i] \geq \tau \\
1, & \text{otherwise}
\end{cases}
\]

where $\tau$ is a threshold that determines which features are considered high perturb-ability. Unlike binary feature mask optimization, which applies the mask through Hadamard product (element-wise multiplication), our approach replaces masked features with neutral values:

\[ 
x_i^{\prime} =
\begin{cases}
\nu_i, & \text{if } m_i = 0 \\
x_i, & \text{otherwise}
\end{cases}
\]

where $x_i$ is the original feature value, $x_i^{\prime}$ is the masked feature value, and $\nu_i$ is a neutral value for feature $i$ (typically the mean or median of that feature from the training data).

\begin{comment}

Algorithm \ref{algo} outlines the feature masking procedure for high PS features.

\begin{algorithm}
\caption{Feature Masking for High PS Features}
\label{algo}
\begin{algorithmic}[1]
\Require Feature vector $\mathbf{x} \in \mathbb{R}^M$, PS scores $\text{PS}{\text{Total}}[f_i]$ for each feature $i \in {1,2,...,M}$, threshold $\tau$, neutral values $\nu_i$ for each feature $i$
\Ensure Masked feature vector $\mathbf{x}^{\prime}$
\State Initialize $\mathbf{x}^{\prime} \leftarrow \mathbf{x}$
\For{$i = 1$ to $M$}
\If{$\text{PS}{\text{Total}}[f_i] \geq \tau$}
\State $x^{\prime}_i \leftarrow \nu_i$ \Comment{Replace high PS feature with neutral value}
\EndIf
\EndFor
\State \Return $\mathbf{x}^{\prime}$
\end{algorithmic}
\end{algorithm}

\end{comment}

By using neutral values instead of always zeros, our approach aligns with feature amputation techniques, which have been shown to be more effective in maintaining model performance when dealing with missing or corrupted features. This is particularly important in the context of adversarial defense, where we aim to minimize the impact on legitimate traffic classification while maximizing robustness against attacks. Moreover, replacing with neutral values maintains the overall statistical distribution of the data, reducing the risk of introducing artifacts that could affect model performance. The neutral values $\nu_i$ could be a single value for all features (for example, 0.5 if the model uses min-max normalization or 0 if the model uses standardization), or it can be a designated value per feature computed from \textbf{the training data} (e.g., mean or median) and stored for use during inference. In our experiments, we tested multiple values of $\nu_i$ to validate our work.

\textbf{For our Option B defense, we test two variants: B1 and B2. Option B1 integrates feature masking during both training and inference}, selectively suppressing high-perturbability features while preserving the original model architecture and input dimensionality. This eliminates the need for pipeline restructuring or retraining on reduced feature subsets (as required by Option A), ensuring compatibility with deployed systems. \textbf{Option B2, in contrast, applies masking exclusively during inference} by replacing high-PS features with precomputed neutral values (e.g., training-data means or medians). This inference-only approach incurs negligible computational overhead, requires no retraining, and enables real-time adaptation to evolving adversarial threats by adjusting masked features on-the-fly. While B1 ensures robustness through training-phase integration, B2 prioritizes operational agility, as the system can dynamically adjust which features to mask based on evolving threat landscapes without requiring retraining or modifications to pretrained models. Our PS-enabled feature masking techniques reduce attack surface: By replacing easily perturbable features with neutral values, we effectively remove attackers' ability to manipulate those features while maintaining the model's structure.

\section{Experimental Environment}
\label{EE}
\begin{figure}
\centering
\captionsetup[subfloat]{labelfont=scriptsize,textfont=scriptsize}
\centering
\subfloat[The correlation Matrix\\ of UNSW-NB15]{%
\includegraphics[width=0.5\linewidth, keepaspectratio=true]{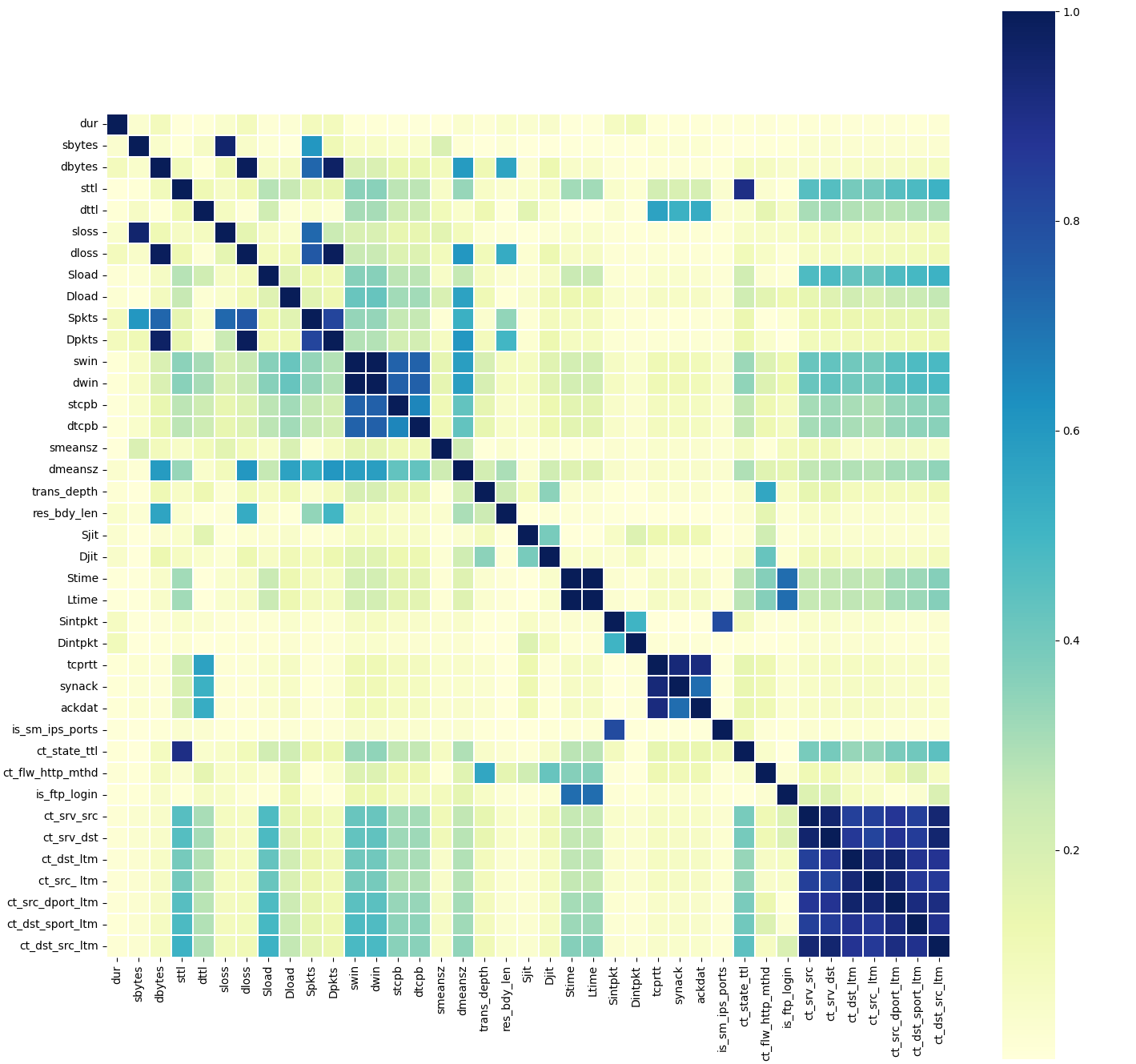}
\label{fig:CORRM15}
}
\subfloat[The correlation Matrix\\ of Improved CIC-IDS2018]{%]{%
\includegraphics[width=0.5\linewidth, keepaspectratio=true]{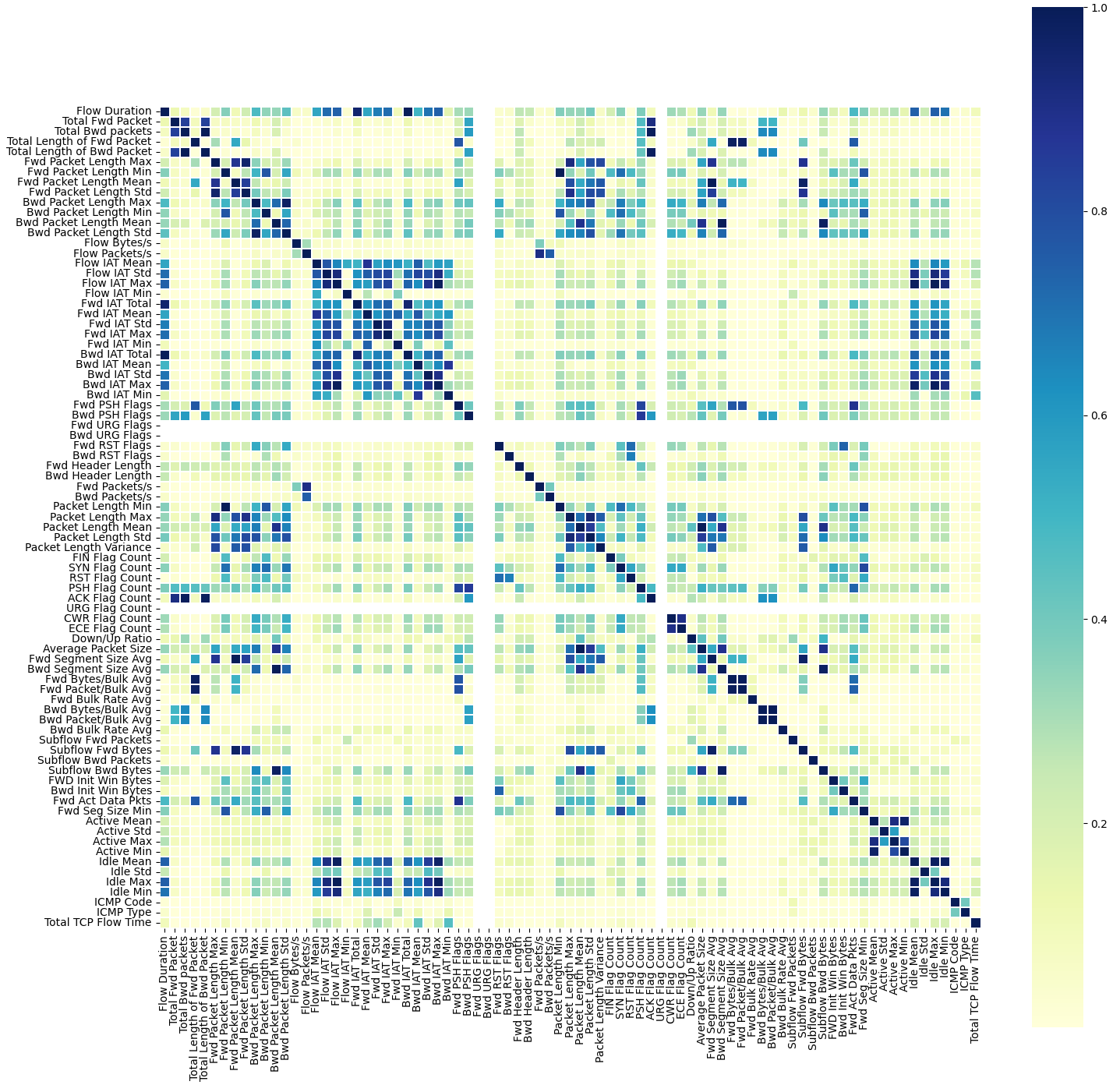}
\label{fig:CORRMCSE}
}
\caption{The correlation Matrices between the features in the used dataset, where darker colors mean higher correlation}
\label{fig:CORRM}
\end{figure}

\begin{figure}
\centering
\captionsetup[subfloat]{labelfont=scriptsize,textfont=scriptsize}
\subfloat[{\scriptsize The correlated features in UNSW-NB15 Dataset}]{%
    \includegraphics[width=0.45\linewidth,keepaspectratio=true]{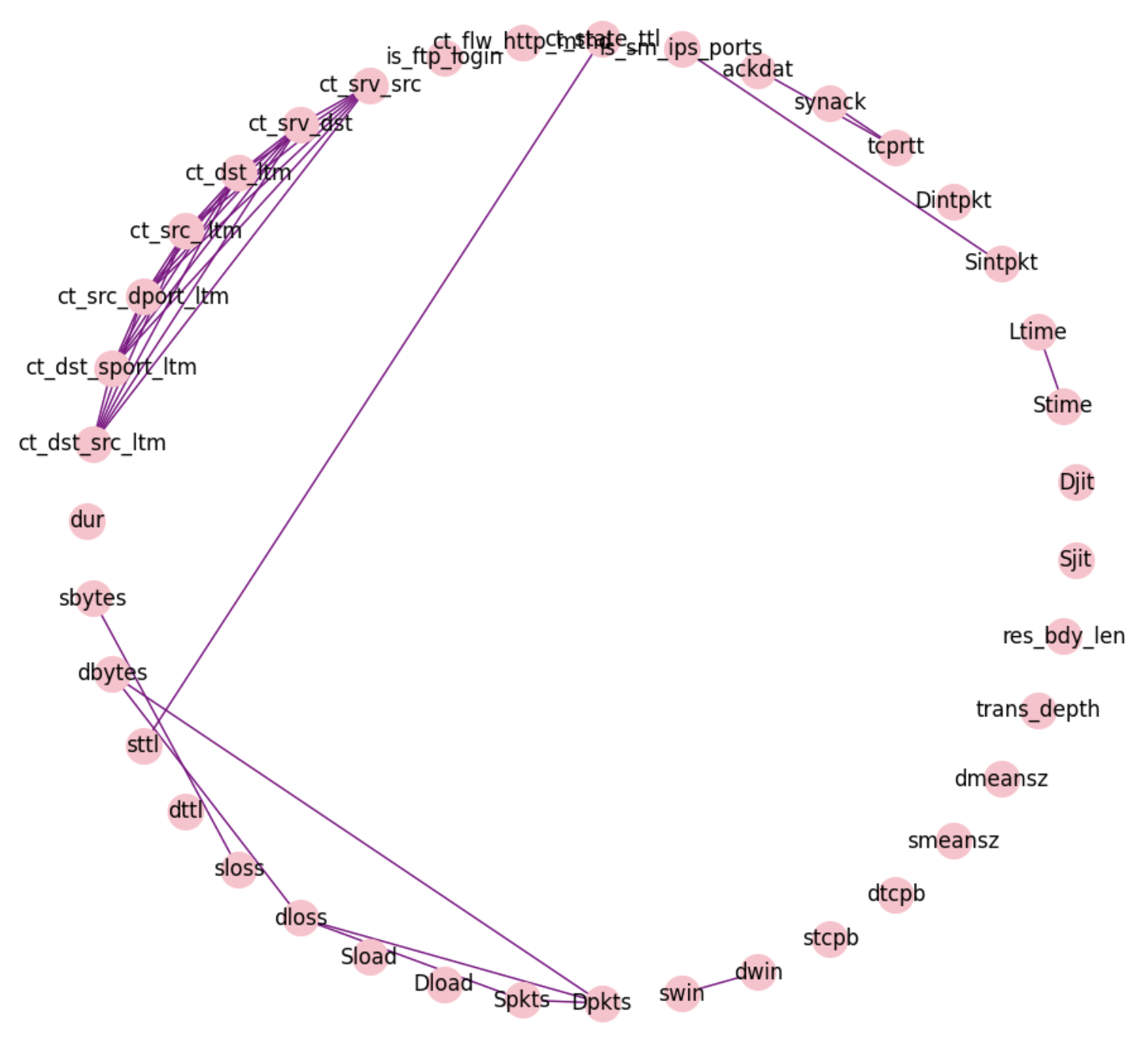}
    \label{fig:CorrUNSW}
}%
\hfill
\subfloat[{\scriptsize The correlated features in Improved CSE-CIC-IDS2018 Dataset}]{%
    \includegraphics[width=0.45\linewidth,keepaspectratio=true]{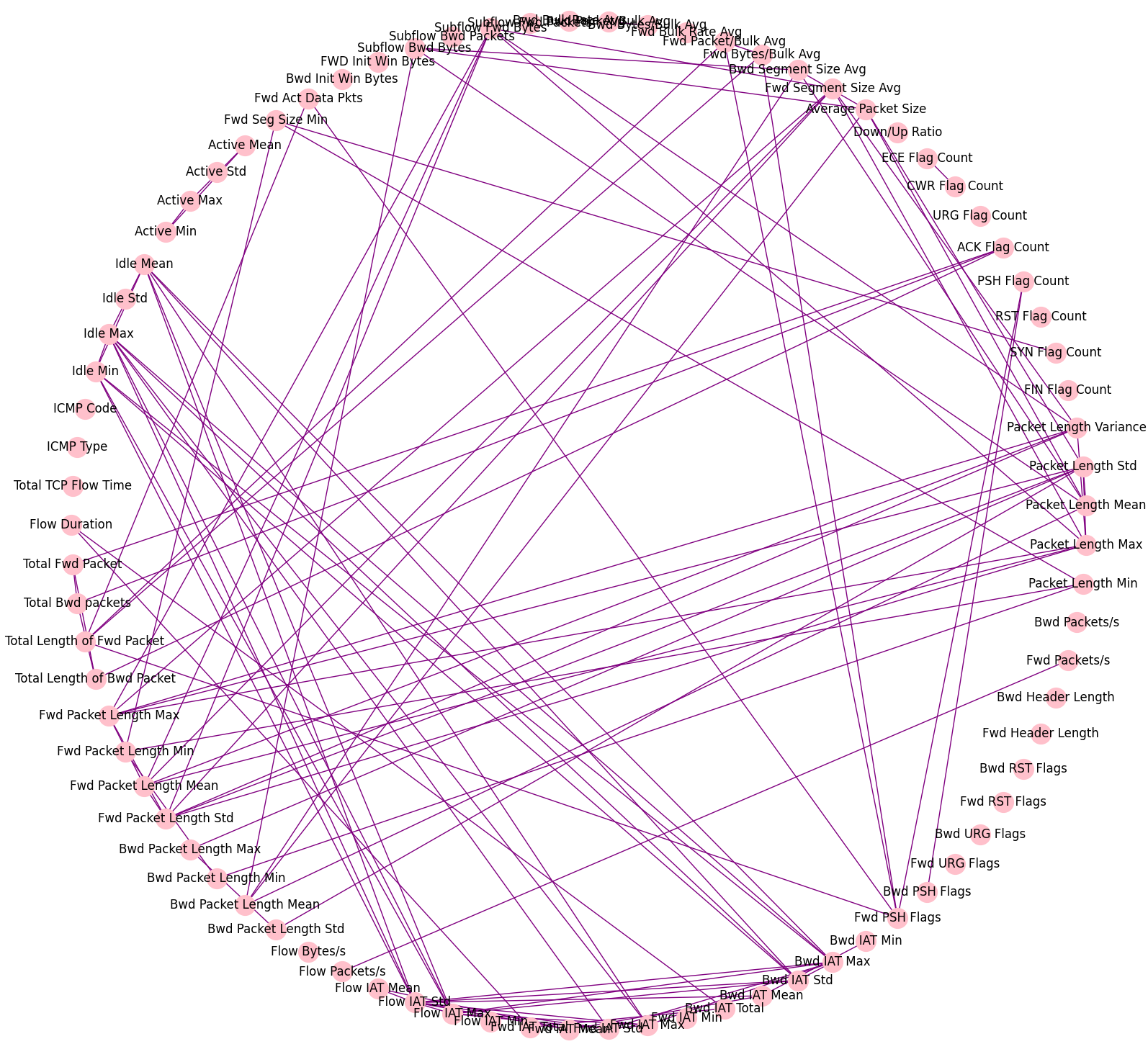}
    \label{fig:CorrCSE}
}
\caption{The correlated features in the used datasets}
\label{fig:Corr}
\end{figure}

\label{ExperimentalEnv}
As seen in Fig. \ref{fig:PSAsaDefense}, the target ML-based NIDS consists of multiple phases. We crafted our own ML-NIDS for our experiments. We will explore our ML-NIDS in the next subsections.

\subsection{Datasets}

For our experiments, we used 3 public NIDS datasets; \textbf{(1) UNSW-NB15 \cite{moustafa2015unsw}}: A widely-used NIDS dataset with 2.5M records (2.2M normal, 0.32M attacks) across 9 attack categories. It includes 49 features and represents realistic network traffic and diverse attack vectors. \textbf{(2) CSE-CIC-IDS2018 (Improved) \cite{liu2022error}}: A dataset with various attacks (e.g., Brute Force, DoS, DDoS) and benign traffic. We use the improved version by Liu et al. \cite{liu2022error}, which resolves issues in the original dataset \cite{sharafaldin2018toward} related to feature generation and labeling, ensuring better reliability for research. \textbf{(3) MCFP \cite{stratospheremalware}}: Unlike the two previous datasets which had CSV feature-space formats, MCFP is a problem-space dataset consisting of PCAP files capturing real network traffic. MCFP was created by the Malware Capture Facility Project at Stratosphere Lab and consists of benign and various malware types' raw network traffic captures in realistic environments, along with their ground-truth information

\subsection{Features' Analysis and Correlations}

As previously explained, some PS fields depend on the analysis of correlations between the dataset's features. This comprehensive analysis begins by examining the dataset's structure and the number of unique values in each column. A correlation matrix is generated to compute the absolute correlations among the remaining features, seen in Fig. \ref{fig:CORRM}.

The correlation matrix employs Pearson's correlation coefficient to measure the linear relationship between pairs of features, resulting in values ranging from -1 to 1. A value of 1 indicates a perfect positive linear correlation, -1 indicates a perfect negative correlation, and 0 indicates no correlation. This matrix captures these relationships, facilitating the identification of highly correlated features.

Moreover, we count and report the number of highly correlated features for each feature using the correlation matrix. This step is crucial for PS\textsubscript{3}[$f_i$] calculations. Additionally, a graph representation of the correlation matrix is constructed using NetworkX \cite{networkx}, where nodes represent features and edges denote strong correlations, as seen in Fig. \ref{fig:Corr}. The graph is visualized to illustrate the interconnectedness of highly correlated features, enhancing the understanding of the dataset's structure, which is crucial for the PS architecture and evaluation.

\subsection{Pre-processing}
\label{Pre}

The pre-processing algorithm prepares a dataset for machine learning by executing several key steps.

%We begin by converting all feature columns to numeric format, replacing non-numeric values with NaN, and removing any rows with missing values. The algorithm separates the features from the target variable and encodes the target using a label encoder if it contains categorical data.
We start by converting all feature columns to numeric format, replacing non-numeric values with NaN, and removing rows with missing values. Identification and non-generalizable features, such as flow ID and timestamps, were also dropped. Our algorithm then separates the features from the target variable, encoding the target with a label encoder if it contains categorical data.

One-hot encoding is applied to categorical features. To maintain high model performance with fewer features, we utilized low Pert. (Green) features to extract useful information, such as the region from the destination IP (using the ipapi Python library \cite{ipapi}) and the application from the destination port number. This information is then one-hot encoded before being fed to the models.

Some researchers, such as Arp et al. \cite{arp2022and}, caution that using the IP address as a feature might lead to spurious correlations or false associations, as the model could learn to identify specific IP ranges instead of recognizing generic attack patterns. However, in our tests, we do not use the IP address as a numerical value. Instead, we extract meaningful geolocation information (e.g., country, region) to provide contextually relevant insights. This approach aims to improve real-world intrusion detection while mitigating the risks associated with direct IP-based correlations. 

Twenty percent of the dataset was allocated for testing. Feature standardization is applied to numerical features, ensuring a mean of zero and a standard deviation of one, enhancing the model's performance. To prevent data leakage, standardization parameters were computed using only the training/validation sets and then applied to the test set. To address class imbalance, we employed random undersampling \cite{liu2020dealing} on the training set to create a balanced dataset.

%Twenty per cent of the dataset was allocated for testing. Feature standardization is applied to numerical features, ensuring a mean of zero and a standard deviation of one, enhancing the model's performance. Standardization was applied after splitting to ensure that information from the test set does not influence the model during training, which would lead to data leakage. To mitigate data leakage, we computed the mean and standard deviation on the training/validation sets and then apply the same transformation to both the training and test sets. To address class imbalance, random undersampling \cite{liu2020dealing} is employed on the training set, resulting in a balanced training dataset. %This preprocessing pipeline produces resampled training data, test data, and the names of the relevant features for further analysis.

%Further details on our models' architectures will be available in the full write-up.

%\begin{itemize}
 %   \item UNSW-NB15 Dataset \cite{moustafa2015unsw}
 %   \item CSE-CIC-IDS2018 Dataset \cite{liu2022error} 
%\end{itemize}
\subsection{Machine and Deep Learning Models}

Several ML models were employed to create different versions of our experimental NIDS, enabling more robust comparisons and exploration. The models include a Vanilla Neural Network, a Support Vector Machine (SVM), an ensemble model (Random Forest), and a Deep Learning (DL) model (Convolutional Neural Network, CNN). A detailed description of each follows.

\textbf{(1) Vanilla Neural Network:} A 3-layer model (64, 32, 16 neurons) with ReLU activation, a sigmoid output layer, optimized with Adam, trained for 10 epochs (batch size 32). \textbf{(2) SVM:} A linear kernel SVM trained on the resampled dataset to maximize the margin between support vectors and minimize classification errors.
\textbf{(3) Random Forest:} An ensemble of 100 trees for binary classification, reducing overfitting and providing feature importance analysis. \textbf{(4) CNN:} A 1D CNN with 64 filters, max-pooling, and a dense layer (100 neurons), ending with a softmax output. Trained for 10 epochs (batch size 32).

\section{Research Questions}

%In this section, we outline our evaluation plan for the PS metric, focusing on several goals, each tied to specific research questions (RQs) that guide our investigation into the effectiveness and applicability of PS in enhancing the robustness of ML-NIDS. 

\begin{table}[]
%\tiny
\notsotiny
%\scriptsize
\centering
\caption{The number and percentage of features in every perturb-ability class, based on our proposed PS, where green indicates low
perturb-ability features class, yellow indicates medium perturb-ability features class, and red indicates high perturb-ability features class}
\begin{tabular} 
{|p{2.33cm}|p{1cm}|p{1cm}|p{1cm}|p{0.4cm}|}
%{|c|c|c|c|c|}
  \hline
\diagbox{\textbf{Dataset}}{\textbf{Pert. Class}} & \cellcolor{mygreen}\makecell{\textbf{ \# and \%} \\ \textbf{of Low}\\\textbf{Pert.}\\\textbf{Features}} & \cellcolor{yellow}\makecell{\textbf{\# and \% }\\ \textbf{of Med.}\\ \textbf{Pert.}\\ \textbf{Features}} & \cellcolor{myred}\makecell{\textbf{\# and \% }\\ \textbf{of High}\\ \textbf{Pert.}\\ \textbf{Features}}& \textbf{Total}\\
\hline 
\textbf{UNSW-NB15 \cite{moustafa2015unsw}} & \cellcolor{mygreen}\textbf{25 (53.2\%) }& \cellcolor{yellow}\textbf{4 (8.5\%)} & \cellcolor{myred}\textbf{18 (38.3\%)} &\textbf{47}\\
\hline 
\textbf{CSE-CIC-IDS2018$^{\mathrm{*}}$ \cite{liu2022error}} & \cellcolor{mygreen}\textbf{38 (43.2\%)} & \cellcolor{yellow}\textbf{19 (21.6\%)}&  \cellcolor{myred}\textbf{31 (35.2\%)}& \textbf{88}\\
\hline 
\multicolumn{5}{l}{$^{\mathrm{*}}$ Improved CSE-CIC-IDS2018 Dataset by Liu et al. \cite{liu2022error}}

\end{tabular}

	\label{NUMFeaturesTable}
\end{table} 

\begin{figure}
\centering
\captionsetup[subfloat]{labelfont=scriptsize,textfont=scriptsize}
\subfloat[{\scriptsize UNSW-NB15 Dataset}]{%
    \includegraphics[scale=1, width=0.45\linewidth, height=0.22\linewidth]{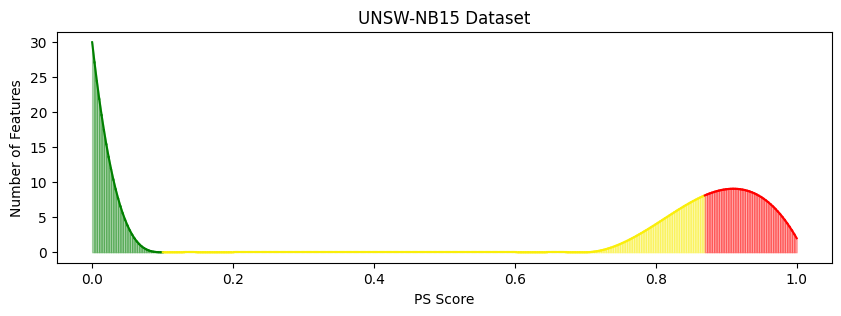}
    \label{fig:rDistUNS}
}%
\hfill
\subfloat[{\scriptsize Improved CSE-CIC-IDS2018 Dataset}]{%
    \includegraphics[scale=1, width=0.45\linewidth, height=0.22\linewidth]{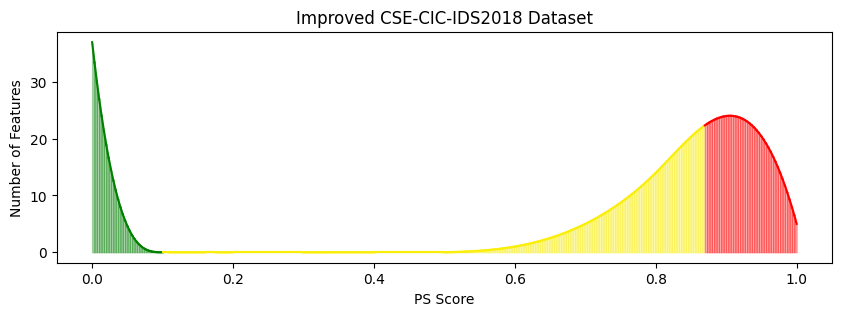}
    \label{fig:rDistCSE}
}
\caption{The histogram of PS values for each dataset where green indicates low perturb-ability features class, yellow indicates medium perturb-ability features class, and red indicates high perturb-ability features class.}
\label{fig:Dist}
\end{figure}

\begin{table*}[]
\notsotiny
\centering
\caption{UNSW-NB15 Dataset's features classified based on our proposed PS, where green indicates a feature with low perturb-ability, yellow indicates a feature with medium perturb-ability, and red indicates a feature with high perturb-ability.}
\begin{tabular}{|c|c|c|c|c|c|c|c|c|c|c|}
  \hline
\cellcolor{mygreen}srcip &
\cellcolor{myred}sport &
\cellcolor{mygreen}dstip &
\cellcolor{mygreen}dsport &
\cellcolor{mygreen}proto &
\cellcolor{mygreen}state &
\cellcolor{myred}dur &
\cellcolor{myred}sbytes &
\cellcolor{yellow}dbytes &
\cellcolor{myred}sttl &
\cellcolor{mygreen}dttl \\
  \hline
\cellcolor{myred}sloss &
\cellcolor{yellow}dloss &
\cellcolor{mygreen}service &
\cellcolor{myred}Sload &
\cellcolor{yellow}Dload &
\cellcolor{myred}Spkts &
\cellcolor{yellow}Dpkts &
\cellcolor{myred}swin &
\cellcolor{myred}dwin &
\cellcolor{myred}stcpb &
\cellcolor{mygreen}dtcpb \\
   \hline
\cellcolor{myred}smeansz &
\cellcolor{mygreen}dmeansz &
\cellcolor{mygreen}trans\_depth &
\cellcolor{mygreen}res\_bdy\_len &
\cellcolor{myred}Sjit &
\cellcolor{mygreen}Djit &
\cellcolor{myred}Stime &
\cellcolor{myred}Ltime &
\cellcolor{myred}Sintpkt &
\cellcolor{mygreen}Dintpkt &
\cellcolor{myred}tcprtt \\
   \hline
\cellcolor{myred}synack &
\cellcolor{myred}ackdat &
\cellcolor{mygreen}is\_sm\_ips\_ports &
\cellcolor{mygreen}ct\_state\_ttl &
\cellcolor{mygreen}ct\_flw\_http\_mthd &
\cellcolor{mygreen}is\_ftp\_login &
\cellcolor{mygreen}ct\_ftp\_cmd &
\cellcolor{mygreen}ct\_srv\_src &
\cellcolor{mygreen}ct\_srv\_dst &
\cellcolor{mygreen}ct\_dst\_ltm &
\cellcolor{mygreen}ct\_src\_ ltm \\
   \hline
\cellcolor{mygreen}ct\_src\_dport\_ltm &
\cellcolor{mygreen}ct\_dst\_sport\_ltm &
\cellcolor{mygreen}ct\_dst\_src\_ltm
   \\
  \cline{1-3}

  % Adjustments made to ensure all features fit into 11 columns.

\end{tabular}
\label{tab:NB15PS2}
\end{table*}

\begin{table*}[]
\notsotiny
\centering
\caption{CSE-CIC-IDS2018 Dataset's features classified based on our proposed PS, where green indicates a feature with low perturb-ability, yellow indicates a feature with medium perturb-ability, and red indicates a feature with high perturb-ability.}
\begin{tabular}{|c|c|c|c|c|c|c|}
  \hline
\cellcolor{mygreen} Flow ID &                  
\cellcolor{mygreen} Src IP &                
\cellcolor{myred} Src Port &                  
\cellcolor{mygreen} Dst IP &                 
\cellcolor{mygreen} Dst Port &  
\cellcolor{mygreen} Protocol & 
\cellcolor{myred} Timestamp
\\
  \hline
\cellcolor{myred} Flow Duration &         
\cellcolor{myred} Total Fwd Packet &        
\cellcolor{yellow} Total Bwd packets & 
\cellcolor{myred} Total Len of Fwd Pack & 
\cellcolor{yellow} Total Len of Bwd Pack &  
\cellcolor{myred} Fwd Packet Length Max &     
\cellcolor{myred} Fwd Packet Length Min
\\
   \hline
\cellcolor{yellow} Fwd Packet Length Mean &  
\cellcolor{yellow} Fwd Packet Length Std & 
\cellcolor{mygreen} Bwd Packet Length Max &   
\cellcolor{mygreen} Bwd Packet Length Min &   
\cellcolor{mygreen} Bwd Packet Length Mean &
\cellcolor{mygreen} Bwd Packet Length Std &
\cellcolor{yellow} Flow Bytes/s                    
\\
   \hline
\cellcolor{yellow} Flow Packets/s &        
\cellcolor{mygreen} Flow IAT Mean &        
\cellcolor{mygreen} Flow IAT Std &        
\cellcolor{mygreen} Flow IAT Max &                 
\cellcolor{mygreen} Flow IAT Min &          
\cellcolor{myred} Fwd IAT Total &          
\cellcolor{yellow} Fwd IAT Mean
\\
   \hline
\cellcolor{yellow} Fwd IAT Std &          
\cellcolor{myred} Fwd IAT Max &  
\cellcolor{myred} Fwd IAT Min &           
\cellcolor{mygreen} Bwd IAT Total &           
\cellcolor{mygreen} Bwd IAT Mean &         
\cellcolor{mygreen} Bwd IAT Std &         
\cellcolor{mygreen} Bwd IAT Max   
\\
   \hline
\cellcolor{mygreen} Bwd IAT Min &            
\cellcolor{myred} Fwd PSH Flags &           
\cellcolor{myred} Bwd PSH Flags &           
\cellcolor{yellow} Fwd URG Flags &           
\cellcolor{mygreen} Bwd URG Flags & 
\cellcolor{yellow} Fwd RST Flags &         
\cellcolor{mygreen} Bwd RST Flags     
\\
   \hline
\cellcolor{myred} Fwd Header Length &         
\cellcolor{mygreen} Bwd Header Length &         
\cellcolor{myred} Fwd Packets/s &  
\cellcolor{mygreen} Bwd Packets/s &       
\cellcolor{myred} Packet Length Min &         
\cellcolor{myred} Packet Length Max &         
\cellcolor{yellow} Packet Length Mean
\\
   \hline
\cellcolor{yellow} Packet Length Std &          
\cellcolor{myred} Packet Len Variance &    
\cellcolor{myred} FIN Flag Count &             
\cellcolor{myred} SYN Flag Count &            
\cellcolor{myred} RST Flag Count &           
\cellcolor{myred} PSH Flag Count &              
\cellcolor{myred} ACK Flag Count
\\
   \hline
\cellcolor{myred} URG Flag Count &          
\cellcolor{myred} CWR Flag Count &          
\cellcolor{myred} ECE Flag Count &         
\cellcolor{yellow} Down/Up Ratio &   
\cellcolor{yellow} Average Packet Size &       
\cellcolor{yellow} Fwd Segment Size Avg &      
\cellcolor{yellow} Bwd Segment Size Avg 
\\
    \hline
\cellcolor{yellow} Fwd Bytes/Bulk Avg &           
\cellcolor{yellow} Fwd Packet/Bulk Avg & 
\cellcolor{myred} Fwd Bulk Rate Avg &        
\cellcolor{mygreen} Bwd Bytes/Bulk Avg &       
\cellcolor{mygreen} Bwd Packet/Bulk Avg &      
\cellcolor{mygreen} Bwd Bulk Rate Avg &     
\cellcolor{myred} Subflow Fwd Packets   
\\
    \hline
\cellcolor{myred} Subflow Fwd Bytes &         
\cellcolor{mygreen} Subflow Bwd Packets &        
\cellcolor{mygreen} Subflow Bwd Bytes &       
\cellcolor{myred} FWD Init Win Bytes &      
\cellcolor{mygreen} Bwd Init Win Bytes &  
\cellcolor{myred} Fwd Act Data Pkts &     
\cellcolor{yellow} Fwd Seg Size Min  
\\
    \hline
\cellcolor{mygreen} Active Mean &            
\cellcolor{mygreen} Active Std &            
\cellcolor{mygreen} Active Max & 
\cellcolor{mygreen} Active Min &              
\cellcolor{mygreen} Idle Mean &            
\cellcolor{mygreen} Idle Std &            
\cellcolor{mygreen} Idle Max
\\
    \hline
\cellcolor{mygreen} Idle Min &  
\cellcolor{mygreen} ICMP Code &                  
\cellcolor{mygreen} ICMP Type &                 
\cellcolor{myred} Total TCP Flow Time 

\\
  \cline{1-4}
\end{tabular}

	\label{tab:CSEPS}
\end{table*}

The following are our goals and their associated Research Questions (RQs) for each:

\begin{enumerate}[label=\Roman*.]

\item \textbf{Knowing the Distribution of Features' PS Classification}
\begin{itemize}
\item \textbf{RQ1:} How are ML-NIDS features characterized and distributed across PS classes, and what criteria determine the classification cutoffs? (\cref{class})
\end{itemize}
\item \textbf{Testing the Possibility of Deploying PS-Enabled Defenses}
\begin{itemize}
\item \textbf{RQ2:} Can ML-NIDS models utilizing only low perturb-ability (green) features or a combination of low and medium perturb-ability (green and yellow) features perform as effectively as models that include all features? (Option A defense possibility) (\cref{perf})
\item \textbf{RQ3:} Can ML-NIDS models with high perturb-ability (red) features and/or medium perturb-ability (yellow) features masked perform as effectively as models that maintain all features unmasked? (Option B defenses possibility) (\cref{perf})
\end{itemize}

\item \textbf{Validation of PS as an indicator to perturb-able features}
\begin{itemize}
\item \textbf{RQ4:} Does PS effectively identify the easily perturb-able features exploited in problem-space adversarial attacks against NIDS documented in research? (\cref{map})
\end{itemize}

\item \textbf{Testing the Effect of PS-Enabled Defenses on Adversarial Attacks}
\begin{itemize}
\item \textbf{RQ5:} Can our PS-enabled defenses neutralize adversarial attacks against flow-based ML-NIDS? (\cref{advtest})
\end{itemize}

\end{enumerate}

Through this evaluation plan, we aim to comprehensively assess the utility of the Perturb-ability Score metric in enhancing the robustness of ML-based NIDS, contributing to the ongoing efforts to secure network systems against adversarial threats.

\section{Features' Classification Using PS (RQ1)}
\label{class}

Table~\ref{NUMFeaturesTable} presents the distribution of features from the two feature-space datasets (UNSW-NB15 and the improved CSE-CIC-IDS2018) across different perturb-ability classes, based on our proposed perturb-ability Score (PS). Features classified as low perturb-ability (green) have a PS of 0, indicating robustness to adversarial manipulation. High perturb-ability features (red) are characterized by a PS greater than or equal to 0.87, marking them as highly perturb-able. We selected the threshold of 0.87 to define highly perturb-able features, as discussed earlier. This threshold is derived from the geometric mean of the five PS components, ensuring that at most one PS field has a value as low as 0.5. In our PS evaluation criteria, a score of 0.5 signifies a condition that tends to reduce the overall PS score, without entirely nullifying it. Using the 0.87 threshold ensures that a feature with at most one such condition is still classified within the high perturb-ability class, thereby reinforcing the rigor of our feature selection criteria. The remaining features, which are neither in the high nor low perturb-ability classes, fall into the medium perturb-ability class (yellow). Table~\ref{NUMFeaturesTable} also shows the percentages of these features across the perturb-ability classes, as defined by our PS. Approximately 35\% of the features in both datasets belong to the high perturb-ability class (red). Fig.~\ref{fig:Dist} depicts the histogram of PS values for each dataset, with the Y-axis representing the number of features and the X-axis representing the PS score. The distribution of PS scores is illustrated, along with the cutoffs between the classes. Table~\ref{tab:NB15PS2} provides a detailed classification of the UNSW-NB15 dataset features according to their perturb-ability scores. Features such as `dstip` and `service` are categorized as low perturb-ability (green), implying their robustness against adversarial manipulations. In contrast, features like `dur` and `stime` are identified as high perturb-ability (red), indicating their susceptibility to adversarial attacks. Table \ref{tab:CSEPS} presents the feature classification for the improved CSE-CIC-IDS2018 dataset. Medium perturb-ability features (yellow), such as `Flow Bytes/s`, occupy an intermediate position between these extremes. It is important to note that low perturb-ability (green) features do not imply that an attacker cannot modify them. Instead, modifying these features may disrupt the network or interfere with the malicious functionality of an attack, or the attacker may have limited or no access to them. Tables \ref{tab:DesUNSW}, \ref{tab:DesCSEA}, and \ref{tab:DesCSEB} in Appendix \ref{AppB} show the definitions of the features in the used datasets.

It is important to note that in all of our experiments, thorough pre-processing was conducted, as discussed in Section \cref{Pre}, and feature engineering was applied to remove spurious or faulty features. Time-based features like timestamps were dropped, as attacks can happen at any time, making these features non-generalizable. Identification features (e.g., flow ID), while green, were removed for the same reason.

\section{Performance of ML-NIDS using various combinations of features based on their perturb-ability (RQ2,3)}
\label{perf}

\newcolumntype{?}{!{\vrule width 1pt}}

\begin{table*}[t]
%\scriptsize
\notsotiny
\caption{The Baseline Performance of an ANN/Random Forest (RF)/SVM/CNN-based NIDS}
\begin{center}
\begin{tabular}{|p{3.5cm}|p{1.2cm}|p{1cm}|p{1cm}|p{1cm}|p{1cm}?p{1cm}|p{1cm}|p{1cm}|p{1cm}|}
\hline
 &  \textbf{Dataset} $\rightarrow$ & \multicolumn{4}{c?}{\textbf {UNSW-NB15}} & \multicolumn{4}{c|}{\textbf {Improved CSE-CIC-IDS2018}}\\

\cline{2-10}
   &\textbf {Model} $\downarrow$ &  \textbf{Accuracy} &  \textbf{Precision} &  \textbf{Recall} &  \textbf{F1} &  \textbf{Accuracy} &  \textbf{Precision} &  \textbf{Recall} &  \textbf {F1} \\

   %\Xhline{4\arrayrulewidth}
\hline
   \textit{ \textbf {Performance of ML-NIDS}}&  \textbf{ANN}  & 0.9883   & 0.9158  & 0.9991  & 0.9557 & 1.0000  & 0.9997  & 0.9998  &   0.9997  \\

  \cline{2-10}
    
  \textit{ \textbf {with all features
Pert.}}   &  \textbf{SVM} & 0.9879  & 0.9129  & 0.9997  & 0.9543   & 0.9999  & 0.9983  & 1.0000  &  0.9991 \\

  \cline{2-10}
   \textit{  \textbf { {\color{green}(Green} + {\color{myellow} Yellow} + {\color{red} Red)}} }  &  \textbf {RF}  & 0.9897  & 0.9251  & 0.9993  & 0.9607  & 1.0000  & 0.9998  & 1.0000  & 0.9999  \\

  \cline{2-10}
 \textit{ \textbf {selected during features selection}}  &  \textbf{CNN}  &  0.9888  & 0.9201   & 0.9976   &  0.9573  &  1.0000 & 0.9995  & 0.9999 &  0.9997 \\

  \hline
\end{tabular}
\end{center}
\label{tab:resBaseline}
\end{table*}

\begin{table*}[thtp]
%\scriptsize
\notsotiny
\caption{The performance of an ANN/RF/SVM/CNN-based NIDS After PS-enabled Feature Selection (Option A Defense).}
\begin{center}
\begin{tabular}{|p{3.5cm}|p{1.2cm}|p{1cm}|p{1cm}|p{1cm}|p{1cm}?p{1cm}|p{1cm}|p{1cm}|p{1cm}|}
\hline
 &  \textbf{Dataset} $\rightarrow$ & \multicolumn{4}{c?}{\textbf {UNSW-NB15}} & \multicolumn{4}{c|}{\textbf {Improved CSE-CIC-IDS2018}}\\

\cline{2-10}
   &\textbf {Model} $\downarrow$ &  \textbf{Accuracy} &  \textbf{Precision} &  \textbf{Recall} &  \textbf{F1} &  \textbf{Accuracy} &  \textbf{Precision} &  \textbf{Recall} &  \textbf {F1} \\

  \hline
  \textit{\textbf {(a) Performance of ML-NIDS}}&  \textbf{ANN}  &  0.9879 & 0.9129  & 0.9998  & 0.9544  & 1.0000  & 0.9998  & 0.9998  & 0.9998   \\

  \cline{2-10}
    
  \textit{\textbf {with only the low
Pert.} }  &  \textbf{SVM} & 0.9879  & 0.9129  &0.9997   & 0.9543  & 0.9999  &0.9984   & 0.9994  & 0.9989   \\

  \cline{2-10}
     \textit{\textbf {features {\color{green}(Green)} selected}}  &  \textbf {RF}  & 0.9891  & 0.9216  & 0.9986  & 0.9585  &  1.0000 & 0.9997   & 1.0000  &  0.9998 \\

  \cline{2-10}
  \textit{\textbf {during features selection}}  &  \textbf{CNN}  & 0.9879  & 0.9126  & 0.9999  & 0.9543  & 1.0000  &  0.9993 &   0.9999 &  0.9996 \\
   
\Xhline{4\arrayrulewidth}

    \textit{\textbf {(b) Performance of ML-NIDS}}&  \textbf{ANN}  & 0.9879   & 0.9127  & 1.0000  & 0.9543  & 0.9998  & 0.9965  & 1.0000  &  0.9983 \\

  \cline{2-10}
    
   \textit{\textbf {with only the low and med
Pert.}}   &  \textbf{SVM} & 0.9879  & 0.9129  & 0.9997  & 0.9543   &  0.9999 &  0.9982 & 0.9998  & 0.9990  \\

  \cline{2-10}
    \textit{ \textbf {features {\color{green}(Green} + {\color{myellow} Yellow)} }}  &  \textbf {RF}  &  0.9892 & 0.9220  & 0.9987  & 0.9588  & 1.0000  & 0.9998  & 1.0000  &  0.9999 \\

  \cline{2-10}
 \textit{ \textbf {selected during features selection}}  &  \textbf{CNN}  & 0.9879  & 0.9128  & 1.0000  &  0.9544 & 1.0000  &  0.9996 &  1.0000 &0.9998   \\

%\Xhline{4\arrayrulewidth}

  \hline
\end{tabular}
\end{center}
\label{tab:res1}
\end{table*}

\begin{table*}[thtp]
%\scriptsize
\notsotiny
\caption{The performance of an ANN/RF/SVM/CNN-based NIDS after PS-enabled feature masking during training and inference (Option B1 Defense).}
\begin{center}
\begin{tabular}{|p{3.5cm}|p{1.2cm}|p{1cm}|p{1cm}|p{1cm}|p{1cm}?p{1cm}|p{1cm}|p{1cm}|p{1cm}|}
\hline
 &  \textbf{Dataset} $\rightarrow$ & \multicolumn{4}{c?}{\textbf {UNSW-NB15}} & \multicolumn{4}{c|}{\textbf {Improved CSE-CIC-IDS2018}}\\

\cline{2-10}
   &\textbf {Model} $\downarrow$ &  \textbf{Accuracy} &  \textbf{Precision} &  \textbf{Recall} &  \textbf{F1} &  \textbf{Accuracy} &  \textbf{Precision} &  \textbf{Recall} &  \textbf {F1} \\

  \hline
  \textit{\textbf {(a) Performance of ML-NIDS}}&  \textbf{ANN}  &  0.9879 & 0.9128   & 0.9998    & 0.9543   & 1.0000    & 0.9994  & 0.9999   & 0.9996   \\

  \cline{2-10}
    
  \textit{\textbf {with only the low
Pert.} }  &  \textbf{SVM} & 0.9879    & 0.9130   &0.9996    & 0.9543  & 0.9999  &0.9982    & 0.9995   & 0.9988   \\

  \cline{2-10}
     \textit{\textbf {features {\color{green}(Green)} not masked}}  &  \textbf {RF}  & 0.9892   & 0.9225  & 0.9982    & 0.9588   &  1.0000 & 0.9998   & 1.0000   &  0.9999  \\

  \cline{2-10}
  \textit{\textbf {({\color{red}(Red} + {\color{myellow} Yellow)} are masked)}}  &  \textbf{CNN}  & 0.9881   & 0.9142   & 0.9994   & 0.9549   & 1.0000   & 0.9997  &   0.9998  &  0.9998   \\

\Xhline{4\arrayrulewidth}

    \textit{\textbf {(b) Performance of ML-NIDS}}&  \textbf{ANN}  &0.9879   & 0.9128    & 1.0000   & 0.9544    & 1.0000    & 0.9996   & 1.0000   &  0.9998   \\

  \cline{2-10}
    
   \textit{\textbf {with only the low and med
Pert.}}   &  \textbf{SVM} &0.9879   & 0.9130   & 0.9997   & 0.9543   &  0.9999  &  0.9982  & 0.9998   & 0.9990   \\

  \cline{2-10}
    \textit{ \textbf {features {\color{green}(Green} + {\color{myellow} Yellow)} not masked }}  &  \textbf {RF}  &  0.9893 & 0.9233   & 0.9981   & 0.9592   & 1.0000   & 0.9998    & 1.0000   &  0.9999  \\

  \cline{2-10}
 \textit{ \textbf {({\color{red}Red} are masked)}}  &  \textbf{CNN}  & 0.9881    & 0.9140   & 0.9998    &  0.9550   & 1.0000   & 0.9997  &  0.9998  & 0.9998   \\

%\Xhline{4\arrayrulewidth}

  \hline
\end{tabular}
\end{center}
\label{tab:res2A}
\end{table*}

\begin{table*}[thtp]
%\scriptsize
\notsotiny
\caption{The performance of an ANN/RF/SVM/CNN-based NIDS after PS-enabled feature masking during inference (Option B2 Defense).}
\begin{center}
\begin{tabular}{|p{3.5cm}|p{1.2cm}|p{1cm}|p{1cm}|p{1cm}|p{1cm}?p{1cm}|p{1cm}|p{1cm}|p{1cm}|}
\hline
 &  \textbf{Dataset} $\rightarrow$ & \multicolumn{4}{c?}{\textbf {UNSW-NB15}} & \multicolumn{4}{c|}{\textbf {Improved CSE-CIC-IDS2018}}\\

\cline{2-10}
   &\textbf {Model} $\downarrow$ &  \textbf{Accuracy} &  \textbf{Precision} &  \textbf{Recall} &  \textbf{F1} &  \textbf{Accuracy} &  \textbf{Precision} &  \textbf{Recall} &  \textbf {F1} \\

  \hline
  \textit{\textbf {(a) Performance of ML-NIDS}}&  \textbf{ANN}  &  0.9862 & 0.9018  & 0.9999   & 0.9483  & 0.9980   & 0.9684  & 0.9999  & 0.9839   \\

  \cline{2-10}
    
  \textit{\textbf {with only the low
Pert.} }  &  \textbf{SVM} & 0.9872   & 0.9079   &0.9999    & 0.9517   & 0.9988  &0.9806    & 0.9990  & 0.9897    \\

  \cline{2-10}
     \textit{\textbf {features {\color{green}(Green)} not masked}}  &  \textbf {RF}  & 0.9841  & \cellcolor{myazure} 0.8883  & 1.0000   & \cellcolor{myazure}0.9408  &  0.9971 & 0.9998   & \cellcolor{myazure} 0.9513  &  0.9749  \\

  \cline{2-10}
  \textit{\textbf {({\color{red}(Red} + {\color{myellow} Yellow)} are masked)}}  &  \textbf{CNN}  & 0.9876   & 0.9110   & 0.9998   & 0.9533  & 0.9977   &  0.9635 &   0.9999  &  0.9814  \\
   
\Xhline{4\arrayrulewidth}

    \textit{\textbf {(b) Performance of ML-NIDS}}&  \textbf{ANN}  &0.9862   & 0.9018   & 0.9999   & 0.9483   & 0.9987   & 0.9793   & 0.9998   &  0.9895  \\

  \cline{2-10}
    
   \textit{\textbf {with only the low and med
Pert.}}   &  \textbf{SVM} & 0.9880   & 0.9133  & 0.9995  & 0.9545   &  0.9985  &  0.9753  & 0.9996   & 0.9873   \\

  \cline{2-10}
    \textit{ \textbf {features {\color{green}(Green} + {\color{myellow} Yellow)} not masked }}  &  \textbf {RF}  &  0.9868  & 0.9054   & 1.0000   & 0.9504   & 0.9999   & 0.9999   & 0.9985   &  0.9992 \\

  \cline{2-10}
 \textit{ \textbf {({\color{red}Red} are masked)}}  &  \textbf{CNN}  & 0.9879   & 0.9129   & 0.9993   &  0.9541  & 0.9982   &  0.9717  &  1.0000  & 0.9856    \\

%\Xhline{4\arrayrulewidth}

  \hline
\end{tabular}
\end{center}
\label{tab:res2}
\end{table*}

\begin{table*}[t]
%\scriptsize
\notsotiny
\centering

\caption{Mapping problem-space evasion adversarial attacks' traffic morphing techniques to features, the features are colored based on our PS classification.$^{\mathrm{*}}$\\}

\begin{tabular}{|p{5.4cm}|p{4cm}|p{6.4cm}|}
\hline
 \cellcolor{mygray} {\footnotesize \textbf{Problem-space Attack and its Problem-space Morphing Techniques}} &\cellcolor{mygray} {\footnotesize \textbf{Potentially Perturb-ed Features in Feature-space in UNSW-NB15}}  & \cellcolor{mygray} {\footnotesize \textbf{Potentially Perturb-ed  Features in Feature-space in improved CSE-CIC-IDS2018}}  \\
\hline
Han et al. \cite{han2021evaluating} modify the interarrival times of packets in the original traffic, change values to the Time to Live (TTL) field, request to establish connections that are already established (or in the process of being established), and add padding to payloads. \cite{han2021evaluating}& \textcolor{red}{sttl, dur, Sjit, Sintpkt, Sload, Stime, Ltime, tcprtt, synack, ackdat} . \textcolor{red}{sbytes, smeansz, Sload,} \textcolor{myellow}{dbytes, Dload}  .  \textcolor{red}{Spkts,} \textcolor{myellow}{Dpkts}. & \textcolor{red}{Flow Duration, Timestamp,} \textcolor{myellow}{Flow Bytes/s, Flow Packets/s,} \textcolor{red}{Fwd IAT Total}, \textcolor{myellow}{Fwd IAT Mean, Fwd IAT Std}, \textcolor{red}{Fwd IAT Max, Fwd IAT Min, Fwd Packets/s,}  .\textcolor{red}{Total Length of Fwd Packet, Fwd Packet Length Max, Min,} \textcolor{myellow}{Fwd Packet Length Mean, Fwd Packet Length Std}, \textcolor{red}{Fwd Bulk Rate Avg}, \textcolor{myellow}{Fwd Bytes/Bulk Avg, Fwd Segment Size Avg,} \textcolor{red}{Subflow Fwd Bytes , Fwd Act Data Pkts}. \textcolor{red}{Total Fwd Packets, Subflow Fwd Packets,} \textcolor{myellow}{Total Bwd Packets, Subflow Bwd Packets} . \textcolor{red}{Fwd PSH Flag, Bwd PSH Flags,} \textcolor{myellow}{Fwd URG Flags, Fwd RST Flags,} \textcolor{red}{FIN Flag Count, SYN Flag Count, RST Flag Count, PSH Flag Count, ACK Flag Count, URG Flag Coun,  CWR Flag Count, ECE Flag Count}\\
 %dur}, \textcolor{red}{Sjit}, \textcolor{red}{Sintpkt}, \textcolor{red}{Sload}, \textcolor{red}{Stime}, \textcolor{red}{Ltime}, \textcolor{red}{tcprtt}, \textcolor{red}{synack}, \textcolor{red}{ackdat} . \textcolor{red}{sbytes}, \textcolor{red}{smeansz}, \textcolor{myellow}{dbytes, Dload} ,\textcolor{green}{dmeansz} . & \textcolor{red}{Flow Duration, Timestamp,} \textcolor{myellow}{Flow Bytes/s, Flow Packets/s,} \textcolor{red}{Fwd IAT Total}, \textcolor{myellow}{Mean, Std}, \textcolor{red}{Max, Min, Fwd Packets/s,} \textcolor{green}{Bwd Packets/s,} .\textcolor{red}{Total Length of Fwd Packet, Fwd Packet Length Max, Min,} \textcolor{myellow}{Mean, Std}, \textcolor{red}{Fwd Bulk Rate Avg}, \textcolor{myellow}{Fwd Bytes/Bulk Avg, Fwd Segment Size Avg,} \textcolor{red}{Subflow Fwd Bytes , Fwd Act Data Pkts}..\\

%& \cellcolor{mypink}  Altering the \# protocol layer of packets in crafted traffic * & &    \\
%%%%%\textcolor{green}{Bwd Packets/s,} 
\hline

\cellcolor{mygray} Hashemi et al. \cite{hashemi2019towards}  split the original packet payload into multiple packets, modify the timing between packets by either increasing or decreasing the intervals, and inject dummy packets with random lengths, transmission times, and flag settings. \cite{hashemi2019towards} &\cellcolor{mygray}  \textcolor{red}{dur}, \textcolor{red}{Sjit}, \textcolor{red}{Sintpkt}, \textcolor{red}{Sload}, \textcolor{red}{Stime}, \textcolor{red}{Ltime}, \textcolor{red}{tcprtt}, \textcolor{red}{synack}, \textcolor{red}{ackdat} . \textcolor{red}{sbytes}, \textcolor{red}{smeansz}, \textcolor{red}{Sload}, \textcolor{myellow}{dbytes}, \textcolor{myellow}{Dload} \textcolor{red}{Spkts}, \textcolor{myellow}{Dpkts}. &\cellcolor{mygray}\textcolor{red}{Flow Duration, Timestamp,} \textcolor{myellow}{Flow Bytes/s, Flow Packets/s,} \textcolor{red}{Fwd IAT Total}, \textcolor{myellow}{Fwd IAT Mean,Fwd IAT Std}, \textcolor{red}{Fwd IAT Max, Fwd IAT Min, Fwd Packets/s,}  .\textcolor{red}{Total Length of Fwd Packet, Fwd Packet Length Max, Fwd Packet Length Min,} \textcolor{myellow}{Fwd Packet Length Mean, Fwd Packet Length Std}, \textcolor{red}{Fwd Bulk Rate Avg}, \textcolor{myellow}{Fwd Bytes/Bulk Avg, Fwd Segment Size Avg,} \textcolor{red}{Subflow Fwd Bytes , Fwd Act Data Pkts}. \textcolor{red}{Total Fwd Packets, Subflow Fwd Packets,} \textcolor{myellow}{Total Bwd Packets, Subflow Bwd Packets} . \textcolor{red}{Fwd PSH Flag, Bwd PSH Flags,} \textcolor{myellow}{Fwd URG Flags, Fwd RST Flags,} \textcolor{red}{FIN Flag Count, SYN Flag Count, RST Flag Count, PSH Flag Count, ACK Flag Count, URG Flag Coun,  CWR Flag Count, ECE Flag Count}\\

\hline

Vitorino et al.\cite{vitorino2023sok} \cite{vitorino2023towards} \cite{vitorino2022adaptative}  modify various flow attributes such as flow duration, average interarrival time between packets, packet rate (packets per second), average forward packet length, smallest forward segment size, minimum interarrival time between packets, and maximum interarrival time. \cite{vitorino2023sok} \cite{vitorino2023towards} \cite{vitorino2022adaptative}   & \textcolor{red}{dur, Sjit, Sload, sbytes, Spkts, Sintpkt,} \textcolor{red}{smeansz} & \textcolor{red}{Flow Duration, Fwd IAT Total,} \textcolor{myellow}{Fwd IAT} \textcolor{myellow}{Mean, Fwd IAT Std,} \textcolor{red}{Fwd IAT Max,}\textcolor{myellow}{Fwd Packet Length Mean,} \textcolor{red}{Fwd IAT Min, Fwd IAT Max,} \textcolor{myellow}{Flow Bytes/s, Flow Packets/s}  \\

\hline
\cellcolor{mygray} Yan et al. \cite{yan2023automatic} modify length-related features by padding packets with irrelevant characters, increase the packet count by duplicating the request multiple times, and modify time-related features by introducing delays before each packet is transmitted from the client. \cite{yan2023automatic}&\cellcolor{mygray} \textcolor{red}{dur, Sjit, Sintpkt, Sload, Stime, Ltime, tcprtt, synack, ackdat} . \textcolor{red}{sbytes, smeansz, Sload,} \textcolor{myellow}{dbytes, Dload} \textcolor{red}{Spkts,} \textcolor{myellow}{Dpkts}. & \cellcolor{mygray}\textcolor{red}{Flow Duration, Timestamp,} \textcolor{myellow}{Flow Bytes/s, Flow Packets/s,} \textcolor{red}{Fwd IAT Total}, \textcolor{myellow}{Fwd IAT Mean, Fwd IAT Std}, \textcolor{red}{Fwd IAT Max, Fwd IAT Min, Fwd Packets/s,}  .\textcolor{red}{Total Length of Fwd Packet, Fwd Packet Length Max, Fwd Packet Length Min,} \textcolor{myellow}{Fwd Packet Length Mean, Fwd Packet Length Std}, \textcolor{red}{Fwd Bulk Rate Avg}, \textcolor{myellow}{Fwd Bytes/Bulk Avg, Fwd Segment Size Avg,} \textcolor{red}{Subflow Fwd Bytes , Fwd Act Data Pkts}. \textcolor{red}{Total Fwd Packets, Subflow Fwd Packets,} \textcolor{myellow}{Total Bwd Packets, Subflow Bwd Packets} . \textcolor{red}{Fwd PSH Flag, Bwd PSH Flags,} \textcolor{myellow}{Fwd URG Flags, Fwd RST Flags,} \textcolor{red}{FIN Flag Count, SYN Flag Count, RST Flag Count, PSH Flag Count, ACK Flag Count, URG Flag Coun,  CWR Flag Count, ECE Flag Count}\\

%\cellcolor{mygray}   Flow Duration, Timestamp, Flow Bytes/s, Flow Packets/s, Fwd IAT Total, Mean, Std, Max, Min, Fwd Packets/s, Bwd Packets/s, .Total Length of Fwd Packet, Fwd Packet Length Max, Min, Mean, Std:, Flow Bytes/s, Fwd Bulk Rate Avg, Fwd Bytes/Bulk Avg, Fwd Segment Size Avg, Subflow Fwd Bytes , Fwd Act Data Pkts. . Total Fwd Packets, Subflow Fwd Packets, Total Bwd Packets, Subflow Bwd Packets . Fwd PSH Flags, Bwd PSH Flags, Fwd URG Flags, Fwd RST Flags, FIN Flag Count, SYN Flag Count, RST Flag Count, PSH Flag Count, ACK Flag Count, URG Flag Coun,  CWR Flag Count, ECE Flag Count \\

\hline
Homoliak et al. \cite{homoliak2018improving} spread out packets over time, drop or duplicate packets, rearrange their order, and perform payload fragmentation. \cite{homoliak2018improving}&
\textcolor{red}{dur, Sjit, Sintpkt, Sload, Stime, Ltime, tcprtt, synack, ackdat} . \textcolor{red}{sbytes, smeansz, Sload,} \textcolor{myellow}{dbytes, Dload}  \textcolor{red}{Spkts,} \textcolor{myellow}{Dpkts}.\textcolor{red}{sloss}

%dur, Sjit, Sintpkt, Sload, Stime, Ltime, tcprtt, synack, ackdat . sbytes, smeansz, Sload, (dbytes, dmeansz, Dload) . . Spkts, Dpkts. sloss
& \textcolor{red}{Flow Duration, Timestamp,} \textcolor{myellow}{Flow Bytes/s, Flow Packets/s,} \textcolor{red}{Fwd IAT Total}, \textcolor{myellow}{Fwd IAT Total Mean, Fwd IAT Total Std}, \textcolor{red}{Fwd IAT Total Max, Fwd IAT Total Min, Fwd Packets/s,}  .\textcolor{red}{Total Length of Fwd Packet, Fwd Packet Length Max, Fwd Packet Length Min,} \textcolor{myellow}{Fwd Packet Length Mean, Fwd Packet Length Std}, \textcolor{red}{Fwd Bulk Rate Avg}, \textcolor{myellow}{Fwd Bytes/Bulk Avg, Fwd Segment Size Avg,} \textcolor{red}{Subflow Fwd Bytes , Fwd Act Data Pkts}. \textcolor{red}{Total Fwd Packets, Subflow Fwd Packets,} \textcolor{myellow}{Total Bwd Packets, Subflow Bwd Packets} \\
%Flow Duration, Timestamp, Flow Bytes/s, Flow Packets/s, Fwd IAT Total, Mean, Std, Max, Min, Fwd Packets/s, Bwd Packets/s,  .Total Length of Fwd Packet, Fwd Packet Length Max, Min, Mean, Std:, Flow Bytes/s, Fwd Bulk Rate Avg, Fwd Bytes/Bulk Avg, Fwd Segment Size Avg, Subflow Fwd Bytes , Fwd Act Data Pkts. . Total Fwd Packets, Subflow Fwd Packets, Total Bwd Packets, Subflow Bwd Packets . \\

\hline

\cellcolor{mygray}Apruzzese et al. \cite{apruzzese2024adversarial} morph features related to data transmission by padding UDP packets, and target only TCP packets with the PSH flag by adding small padding to them and repeating the process. \cite{apruzzese2024adversarial}&\cellcolor{mygray}  \textcolor{red}{sbytes, smeansz, Sload,} \textcolor{myellow}{dbytes, Dload}  & \cellcolor{mygray} \textcolor{red}{Total Length of Fwd Packet, Fwd Packet Length Max, Fwd Packet Length Min,} \textcolor{myellow}{Fwd Packet Length Mean, Fwd Packet Length Std, Flow Bytes/s,} \textcolor{red}{Fwd Bulk Rate Avg,} \textcolor{myellow}{Fwd Bytes/Bulk Avg, Fwd Segment Size Avg,} \textcolor{red}{Subflow Fwd Bytes , Fwd Act Data Pkts} \\

\hline
\multicolumn{3}{l}{$^{\mathrm{*}}$Note: The generation of this table is based on our domain knowledge and our understanding of the writings published by the referenced researchers.}

\end{tabular}

\label{tab:example}
\end{table*}

To evaluate the validity of our PS-enabled defenses, we conducted extensive experiments using four different machine learning models (ANN, SVM, Random Forest (RF), and CNN) on two feature-space datasets (UNSW-NB15 \cite{moustafa2015unsw} and Improved CSE-CIC-IDS2018 \cite{liu2022error}). These tests aim to check if selecting non-perturb-able features or masking all but non-perturb-able features will significantly reduce the performance of the models during our PS-enabled defenses.

%To evaluate the validity of our PS-enabled defenses, we conducted extensive experiments using four different machine learning models (ANN, SVM, Random Forest, and CNN) on two datasets (UNSW-NB15 and Improved CSE-CIC-IDS2018). Our tests 

%answer RQ2 and RQ3, we conducted extensive experiments using four different machine learning models (ANN, SVM, Random Forest, and CNN) on two datasets (UNSW-NB15 and Improved CSE-CIC-IDS2018). The results are presented in Tables \ref{tab:resBaseline}, \ref{tab:res}, and \ref{tab:res2}.

Table \ref{tab:resBaseline} shows the baseline performance of the ML-NIDS models using all features (Green + Yellow + Red). This serves as our reference point for comparison. As we can observe, all models achieve high performance across both datasets, with accuracy, precision, recall, and F1 scores consistently above 0.99 for the Improved CSE-CIC-IDS2018 dataset and above 0.95 for the UNSW-NB15 dataset.

To address RQ2, we evaluated the performance of ML-NIDS models using only low perturb-ability (Green) features and a combination of low and medium perturb-ability (Green + Yellow) features. The results are presented in Table \ref{tab:res1}. Remarkably, we observe that the performance of the models using only Green features (Table \ref{tab:res1}a) or Green + Yellow features (Table \ref{tab:res1}b) is nearly identical to the baseline performance using all features. For instance, the ANN model on the UNSW-NB15 dataset maintains an accuracy of 0.9879 and an F1 score of 0.9544 when using only Green features, compared to 0.9883 and 0.9557 respectively when using all features. This trend is consistent across all models and both datasets, demonstrating that ML-NIDS models can perform effectively using only low or low-medium perturb-ability features. Thus, based on these results, implementing PS-enabled option A defense with only the green features selected may be preferable as it further reduces the attack surface without model performance degradation.

%To address RQ3, we implemented PS-enabled feature masking and evaluated the performance of ML-NIDS models with high perturb-ability (Red) features and/or medium perturb-ability (Yellow) features masked. The results are shown in Table \ref{tab:res2}. When masking both Red and Yellow features (Table \ref{tab:res2}a), we observe a slight decrease in performance of RF compared to the baseline, as seen in the pink cells. However, the models still maintain high performance, with most metrics remaining above 0.90. When masking only Red features (Table \ref{tab:res2}b), the performance is very close to the baseline, with only minimal decreases in some metrics. Thus, masking only Red features might occasionally be the better option, but that depends on the model and environment.

To address RQ3, we evaluated PS-enabled feature masking under two scenarios: masking during both training and inference (Table \ref{tab:res2A}) and masking only during inference (Table \ref{tab:res2}). For both options, we used the mean value per feature from the training sets as the masking value for these experiments. When masking high perturb-ability features (Red/Red+Yellow) during both phases (Table \ref{tab:res2A}), like Table \ref{tab:res1}, performance remains nearly identical to the baseline across all models and datasets, for example, the ANN model on UNSW-NB15 retains 0.9879 accuracy and 0.9543 F1 score when masking Red+Yellow, matching the baseline’s 0.9883 accuracy and 0.9557 F1. This demonstrates that integrating masking into training preserves model efficacy without architectural changes. In contrast, inference-only masking (Table \ref{tab:res2}) shows a slight performance dip, as seen in the light blue cells, but maintains robustness ($>$ 0.94 F1) while enabling real-time adaptability. The minimal degradation in both scenarios validates that masking high-PS features, whether during training or dynamically at inference, effectively hardens models without compromising utility. Option B1’s training-integrated masking ensures stability, while B2’s inference-only approach prioritizes operational flexibility, allowing threat-responsive adjustments without retraining.

\section{Mapping Traffic Morphing of problem-space attacks to ML-NIDS Features (RQ4)}
\label{map}

To validate our proposed PS, we analyzed problem-space evasion adversarial attacks targeting ML-NIDS \cite{han2021evaluating, hashemi2019towards, vitorino2023sok, vitorino2023towards, vitorino2022adaptative, yan2023automatic, homoliak2018improving, apruzzese2024adversarial}. Table \ref{tab:example} presents a detailed mapping of traffic morphing techniques employed in these problem-space adversarial attacks to the corresponding perturbed features within the feature-space of the two datasets utilized in this study. The table outlines the problem-space morphing techniques and maps them to the potentially impacted features. Additionally, it classifies these features according to our perturb-ability Score (PS) system, using a color-coded scheme: low perturb-ability (green), medium perturb-ability (yellow), and high perturb-ability (red).

In problem-space attacks, the attackers use different methods to morph traffic, which acts like adding perturbations to certain features after feature extraction. For instance, modifying the packet lengths by adding padding affects length-related features, such as the ``Total Length of Forward Packets'' in the CSE-CIC-IDS2018 dataset. Another example includes increasing the number of packets or altering time-related features, which impacts metrics like Flow Duration, Total Fwd Packets, and Fwd Inter-Arrival Time (IAT).
%As shown in Table \ref{tab:example}, the majority of perturbed or affected features in these problem-space evasion attacks exhibit high perturb-ability scores, classified under the red category. A smaller subset of features shows medium perturb-ability. This highlights the efficacy of our defense mechanisms, which strategically eliminate or mask high-perturb-ability features, rendering these types of attacks ineffective. Since the attacker manipulates features that are excluded from the final feature set. In other words, our defense will eliminate these easily perturbed features, \textbf{which may render these types of attacks ineffective, as the attacker is modifying features that are excluded from the selection, making their changes irrelevant to the feature vector.} 

As shown in Table \ref{tab:example}, the majority of perturbed or affected features in these problem-space evasion attacks exhibit high perturb-ability scores, classified under the red category. A smaller subset of features shows medium perturb-ability. This highlights the efficacy of our defense mechanisms, which strategically eliminate or mask high-perturb-ability features. \textbf{These defenses significantly reduce the effectiveness of such attacks, as the attacker manipulates features that are either masked or excluded entirely from the final feature set, making their modifications irrelevant to the feature vector used for classification.}

Table \ref{tab:example} shows that limited changes in the problem-space can cause significant, widespread changes in the feature-space. While this may seem to benefit attackers, such manipulations often result in unintended side effects \cite{he2023adversarial}, which we term \textbf{collateral damage}. We coined the term ``collateral damage features'' because ``side effect,'' commonly used in research, is primarily associated with the unintended consequences of medicines, which are generally beneficial. In contrast, within the context of adversarial attacks, these features are unintentionally perturbed by attackers, and adversarial attacks are generally harmful. Therefore, we believe that "collateral damage features" is a more accurate and fitting term. For example, modifying one feature (such as maximum forward IAT) to evade detection may unintentionally alter other correlated features (such as flow duration, packets per second, total forward IAT, and forward IAT mean) due to feature interdependencies in ML-based NIDS. These collateral changes do not follow any particular gradient direction \cite{shehaby2023adversarial}, making their effects unpredictable and potentially undermining the attack.

\section{Adversarial Testing (RQ5)}
\label{advtest}

\subsection{Systematization of ML-NIDS Adversarial Perturbations}

\begin{figure}[t]
\centering
\includegraphics[width=1\linewidth,keepaspectratio=true]{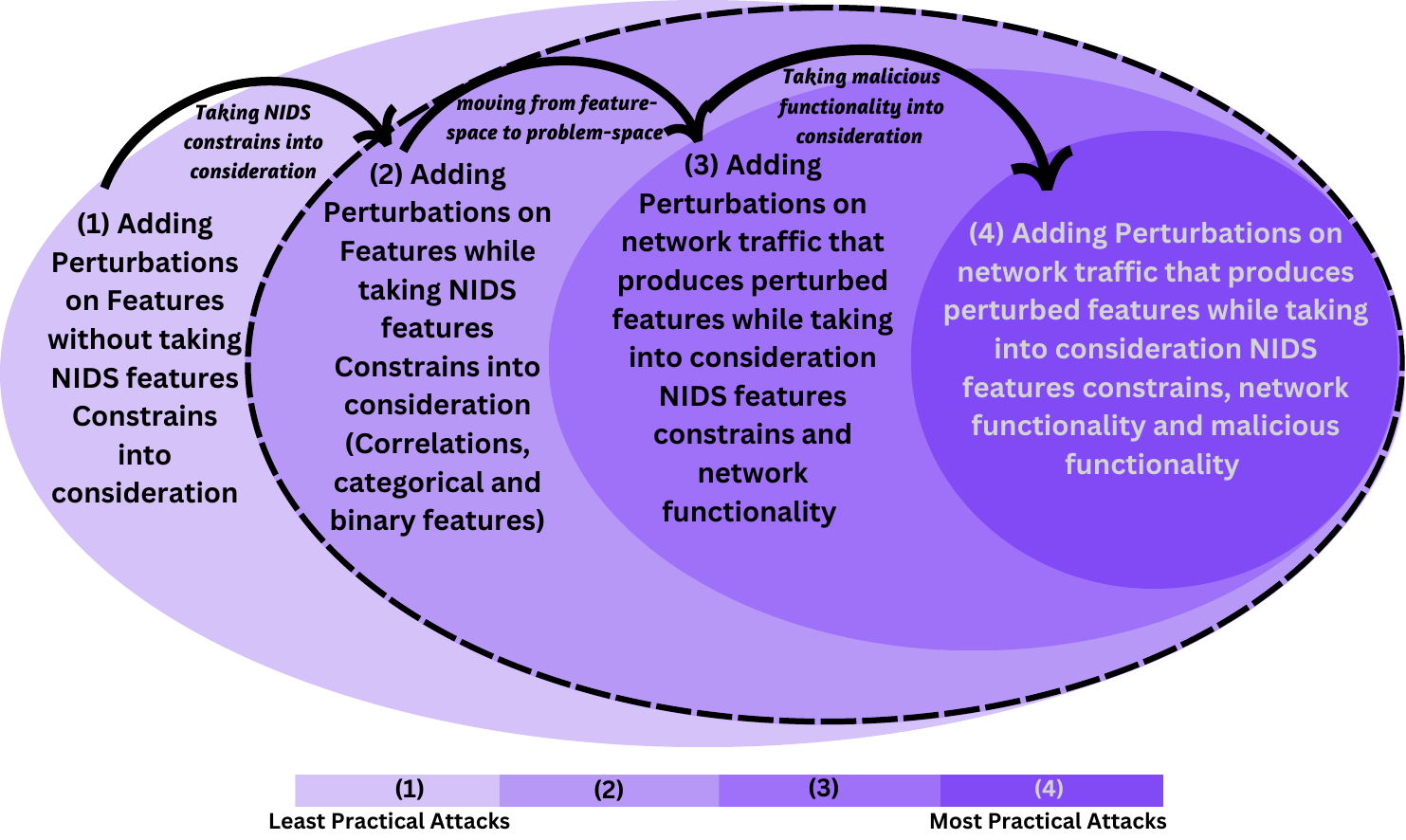}
	\caption{ML-NIDS Adversarial Perturbations Types}
	\label{fig:Circles}
	\centering
\end{figure}

Figure \ref{fig:Circles} presents a hierarchical classification of perturbation attack strategies against ML-NIDS, organized in four levels of increasing practicality and sophistication, with the innermost circle (Level 4) being the most practical and complex to create.
The diagram uses overlapping circles to illustrate how each level builds upon and encompasses the constraints of previous levels:

\textbf{Level 1} (Least Practical Attacks): Adding perturbations on features without taking NIDS features constraints into consideration.
\textbf{Level 2}: Adding perturbations on features while taking NIDS features constraints (correlations, categorical and binary features) into consideration.
\textbf{Level 3}: Adding perturbations on network traffic that produces perturbed features while taking into consideration NIDS features constraints and network functionality, but without maintaining the malicious aim of the attack after altering the network traffic. For example, adding huge delays between packets as perturbations in a A denial of service (DoS) attack.
\textbf{Level 4} (Most Practical Attacks): Adding perturbations on network traffic that produces perturbed features while taking into consideration NIDS features constraints, network functionality and malicious functionality.

%We recommend that attack testing should be conducted at Level 4, as this represents the most realistic adversarial scenario that accounts for all practical constraints an attacker would face.

%For defense testing, we recommend implementing defenses at Level 2, as this provides a balance between practicality and coverage. The nested circular design of our illustration demonstrates how defending at a certain level will implicitly defend against higher levels. 

%Defending at Level 1 is not recommended, as perturbations at this level are extremely impractical and do not require defense in flow-based ML-NIDS.
%For example, implementing defenses at Level 2 will implicitly provide protection against attacks at Levels 3 and 4, since the higher-level attacks must still satisfy the constraints of lower-level attacks. For example, stopping a level 1 or 2 attack will definitely stop a level 3 or 4 attack as problem-space attacks are perturbations on network traffic aiming at producing feature-space perturbations, so if feature-space perturbations don't work then problem-space perturbations will definitely not work.

In our tests to evaluate the effectiveness of our PS-enabled defenses, we utilized Levels 2 and 4, one in the feature-space and the other in the problem-space, to assess the robustness and diversity of our defenses. Moreover, level 2 provides a balance between practicality and coverage. The nested circular design of our illustration demonstrates how defending at a certain level will implicitly defend against higher levels.

%The nested circular design of our illustration demonstrates how defending at a certain level will implicitly defend against higher levels , for example, implementing defenses at Level 2 will implicitly provide protection against attacks at Levels 3 and 4, since the higher-level attacks must still satisfy the constraints of lower-level attacks as problem-space attacks are perturbations on netweok traffic aiming at producing feature-space perturbations.

%\subsubsection{Attack Success Rate (ASR)}
\noindent{\bf Attack Success Rate (ASR):}
We evaluated the impact of adversarial attacks using the Attack Success Rate (ASR), defined as:

\[
\text{ASR} = \frac{\text{Number of Successful Adversarial Attempts}}{\text{Total Number of Adversarial Attempts}} \times 100\%
\]

Prior research frequently defines a successful adversarial attack as occurring when the model's prediction for a perturbed sample diverges from its original output:
\[
f(x_i^*) \neq y_i
\]

Where \( f(x_i^*) \) represents the prediction for the perturbed sample \( x_i^* \) and \( y_i \) denotes the correct prediction. However, we argue this definition combines two distinct scenarios of prediction errors: (1) errors caused by adversarial perturbations and (2) inherent model inaccuracies on unperturbed inputs (the case when even the unperturbed sample produces wrong predictions).

To isolate the adversarial effect, we define a successful attack \textit{only} when perturbations alter the model's original correct prediction ($f(x_i) = y_i$). Our revised ASR formulation reflects this distinction:

%\[
%\text{ASR} = \frac{\sum_{i=1}^{N} 1(f(x_i) \neq f(x_i^*))}{N} \times 100\%
%\]

\[
\text{ASR} = \frac{\sum_{i=1}^{N} 1((f(x_i) = y_i) \land (f(x_i) \neq f(x_i^*)))}{N} \times 100%
\]
where \( f(x_i) \) denotes the model's prediction for original input \( x_i \), \( f(x_i^*) \) represents the perturbed input prediction, \( N \) is the total number of adversarial attempts, and \( 1(\cdot) \) indicates the indicator function (1 if condition holds, 0 otherwise). The condition ($((f(x_i) = y_i) \land (f(x_i) \neq f(x_i^*)))$) ensures we only count instances where the original prediction was correct and the adversarial example successfully changed the prediction.

\subsection{\textbf{Problem-space} Attack vs Our Option A, B1 \& B2 PS-enabled Defenses (level 4)}

%\textcolor{red}{To test if PS effectively captures the easily perturb-able features exploited in problem-space adversarial attacks against flow-based ML-NIDS,}

We conducted a series of experiments using the problem-space blind adversarial perturbation attacks proposed by Apruzzese et al. \cite{apruzzese2024adversarial}. In our opinion, this is one of the most practical adversarial attacks in research. The authors assume that the attacker has no knowledge of the model, cannot query the model, and that their attack operates in problem-space (perturbing network flows) while maintaining NIDS constraints. Their blind perturbations manipulate UDP and TCP packets by adding a small padding of random bytes to their payloads. While applying these manipulations, they ensure that each packet does not exceed its maximum length and recreate the checksum. In our experiment, we used their published attack codes, the same PCAP dataset (MCFP dataset \cite{stratospheremalware}) and the same tool to extract NetFlows from our raw PCAP traces (Argus \cite{argusnetflow}), with the same configuration as Apruzzese et al. \cite{apruzzese2024adversarial}. We created four ML models for classification tasks and performed preprocessing to prepare the data for training.

To evaluate the impact of PS-guided feature selection and masking on the problem-space MCFP dataset, we extended our experiments from Section \ref{perf}. Table \ref{tab:performance_metrics} presents the performance metrics (Accuracy, Precision, Recall, F1-score) for four models (ANN, SVM, RF, CNN) under different defenses: pre-PS (baseline), post-PS feature selection (Option A), post-PS feature masking during training and inference (Option B1), and post-PS feature masking during inference only (Option B2). For both B1 and B2, we used the median value per feature from the training sets as the masking value for these experiments. Both Option A and Option B1 defenses preserved model performance. However, Option B2 resulted in noticeable performance drops in some models, particularly for SVM (cyan cells), and slight reductions in precision for RF and CNN (light blue cells). We will further discuss the limitations and ways to address them for Option B2 in Section \ref{limitations}. For all experiments in this section, IP addresses were excluded during pre-processing to align with Apruzzese et al. \cite{apruzzese2024adversarial}, ensuring consistency.

To test the effectiveness of our PS-enabled defenses against the problem-space blind adversarial perturbation, the features were classified using PS to identify and exclude/mask easily perturbable features. The models were evaluated before and after applying PS-guided defenses. The results, shown in Table \ref{tab:afr_results}, indicate that adversarial attacks by Apruzzese et al., while inconsistent, can reduce ML-NIDS performance. However, after applying PS-guided defenses (our Option A, B1, and B2), these attacks have no measurable effect on model performance. \textbf{The results in Tables \ref{tab:performance_metrics} and \ref{tab:afr_results} confirm that PS-guided defenses effectively mitigate adversarial attacks without compromising performance, except for inference-only masking in Option B2.}

%We conducted a series of experiments using the problem-space blind adversarial attacks by Apruzzese et al \cite{apruzzese2024adversarial}. In our opinion, this is one of the most practical adversarial attacks in research. The authors assume that the attacker has no knowledge about the model, cannot query the model, their attack is in problem-space (perturbing network flows) while maintaining NIDS constraints. Their blind perturbations manipulate UDP and TCP packets by adding a small padding of random bytes to their payload. While applying the manipulations, they ensure that each packet does not exceed its maximum length and recreate the checksum.
%In our experiment, we used the same PCAP dataset (MCFP dataset \cite{stratospheremalware}) and the same tool to extract the NetFlows from our raw PCAP traces (Argus \cite{argusnetflow}) with the same configuration as Apruzzese et al \cite{apruzzese2024adversarial}. We created four ML models for classification tasks and performed preprocessing to prepare the data for training. The features were classified using PS-guided feature selection to identify and exclude easily perturb-able features. The models were evaluated under two conditions: Before and After applying PS-guided feature selection.

\begin{table}[t]
\notsotiny
\caption{Performance metrics (Accuracy, Precision, Recall, F1-score) for models before and after PS feature selection on MCFP Dataset\\}
\centering
\renewcommand{\arraystretch}{1.2}
\begin{tabular}{|l|l|c|c|c|c|}
\hline
\rowcolor[HTML]{EFEFEF} 
\textbf{}                           & \textbf{Model} & \textbf{Acc} & \textbf{Pre} & \textbf{Rec} & \textbf{F1} \\ \hline
\multirow{4}{*}{Pre-PS (Baseline)} 
                                    & ANN            & 0.9998       & 1.0000       & 0.9990       & 0.9995      \\ \cline{2-6} 
                                    & SVM            & 0.9950       & 0.9736       & 0.9970       & 0.9852      \\ \cline{2-6} 
                                    & RF             & 1.0000       & 1.0000       & 1.0000       & 1.0000      \\ \cline{2-6} 
                                    & CNN            & 0.9997       & 0.9980       & 1.0000       & 0.9990      \\ \Xhline{4\arrayrulewidth}
\multirow{4}{*}{\parbox{3.3cm}{Post-PS feature selection\\(Only {\color{green}(Green)}  features Selected)\\Option A Defense}} 
                                    & ANN            & 0.9997       & 0.9990       & 0.9990       & 0.9990      \\ \cline{2-6} 
                                    & SVM            & 0.9963       & 0.9804       & 0.9980       & 0.9891      \\ \cline{2-6} 
                                    & RF             & 1.0000       & 1.0000       & 1.0000       & 1.0000      \\ \cline{2-6} 
                                    & CNN            & 1.0000       & 1.0000       & 1.0000       & 1.0000      \\ 
                                    
                        \Xhline{4\arrayrulewidth}

                                    \multirow{4}{*}{\parbox{3.3cm}{Post-PS feature Masking in Training \\and Inference (Only {\color{green}(Green)}\\features are not masked)\\Option B1 Defense}} 
                                    & ANN            & 0.9998       & 1.0000       & 0.9990       & 0.9995      \\ \cline{2-6} 
                                    & SVM            & 0.9947      & 0.9717       & 0.9970      & 0.9842      \\ \cline{2-6} 
                                    & RF             & 1.0000       & 1.0000       & 1.0000       & 1.0000      \\ \cline{2-6} 
                                    & CNN            & 1.0000       & 1.0000       & 1.0000       & 1.0000      \\            
                                    \Xhline{4\arrayrulewidth}
                                    \multirow{4}{*}{\parbox{3.3cm}{Post-PS feature Masking in Inference\\(Only {\color{green}(Green)} features are not masked)\\Option B2 Defense}} 
                                    & ANN            &   0.9932    &    0.9651    &   0.9950    &   0.9798    \\ \cline{2-6} 
                                    & SVM            & \cellcolor{cyan}   0.8710    &  \cellcolor{cyan} 0.7927     & \cellcolor{cyan}  0.3060   & \cellcolor{cyan} 0.4416     \\ \cline{2-6} 
                                    & RF             &  0.9842      &  \cellcolor{myazure}  0.9132     &   1.0000    & 0.9547     \\ \cline{2-6} 
                                    & CNN            &  0.9792     &  \cellcolor{myazure} 0.8889    &    1.0000    &    0.9412    \\   
                                    
                                    \hline
\end{tabular}

\label{tab:performance_metrics}
\end{table}

\begin{table}[t]
\notsotiny
\caption{ASR Results for TCP and UDP with Pre PS-enabled and Post PS-enabled Defenses (Options A, B1 \& B2) on MCFP Dataset\\}
\centering
\begin{tabular}{|l|l|c|c|}
\hline
\rowcolor[HTML]{EFEFEF} 
\textbf{}                           & \textbf{Model} & \textbf{TCP ASR} & \textbf{UDP ASR} \\ \hline
\multirow{4}{*}{Pre PS-enabled Defenses} 
                                    & ANN            & 0.00\%           & 0.00\%           \\ \cline{2-4} 
                                    & SVM            & 22.50\%  & 39.40\%  \\ \cline{2-4} 
                                    & RF             & 0.00\%           & 0.00\%           \\ \cline{2-4} 
                                    & CNN            & 0.00\%           & 38.40\%  \\ 
                                    \Xhline{4\arrayrulewidth}
                                    %\hline

\multirow{4}{*}{\parbox{4.8cm}{Post \textbf{Option A} Defense: Only {\color{green}(Green)}  features Selected\\\textbf{Options B1 \& B2} showed the exact results of 0\% ASR across all models.\\
In \textbf{Options B1\&2}: only {\color{green}(Green)} features weren't masked.}}
                                    & ANN            & 0.00\%           & 0.00\%           \\ \cline{2-4} 
                                    & SVM            & 0.00\%           & 0.00\%           \\ \cline{2-4} 
                                    & RF             & 0.00\%           & 0.00\%           \\ \cline{2-4} 
                                    & CNN            & 0.00\%           & 0.00\%           \\ \hline

\end{tabular}

\label{tab:afr_results}
\end{table}

\subsection{Constrained \textbf{Feature-space} Attack vs Our Option B2 PS-enabled Defense (level 2)}

To test the effectiveness of our option B2 defense, we conducted a series of experiments using the work by Alhussien et al. \cite{alhussien2024constraining}. They introduced a comprehensive set of network constraints specifically designed to ensure the validity and practicality of adversarial examples in network traffic scenarios. They evaluated their approach using multiple attacks like; Zeroth Order Optimization (ZOO) \cite{chen2017zoo}, a black-box attack relying on gradient estimation through model queries, and Carlini \& Wagner (C\&W) \cite{carlini2017towards}, a white-box optimization-based attack known for producing minimal perturbations. Additionally, Alhussien et al. published portions of their source code, allowing us to utilize and extend their constrained adversarial attack generation methodologies within our experimental evaluation. To reproduce Alhussien et al.'s results and tests, we used the same models, dataset, pre-processing, feature selection, attacks, attack settings, and constraints provided in their published codes. To ensure consistency and avoid altering their architecture or pipeline, we employed Option B2 (inference-only masking), which allowed us to dynamically mask high perturb-ability features during inference while maintaining the original model structure and pre-trained parameters. For more details on Alhussien et al.'s \cite{alhussien2024constraining} constrained attacks and our testing of our defenses against these attacks, including the constraints they imposed on the attacks, the models and attacks we used, and limitations, check Appendix \ref{transWork}.

%the details of these are provided in Table XIII about 

\begin{table}[t]
\notsotiny
\caption{Comparison of Constrained ZOO and C\&W Attacks Pre PS-enabled and Post PS-enabled Feature Masking (Our Option B2 Defense) on UNSW-NB15 Dataset}
\centering
\begin{tabular}{|p{3cm}|p{2cm}|p{2cm}|}

\hline
\cellcolor{mygray} \textbf{Attack} & \cellcolor{mygray} \textbf{Constrained ZOO} & \cellcolor{mygray} \textbf{Constrained C\&W} \\
\hline
\hline
\textbf{Adversarial Test Samples} & 4533 & 4533 \\
\hline
\textbf{Successful Samples} & 897 & 2706 \\
\hline
\textbf{ASR Pre PS-enabled Masking} & 19.79\% & 59.70\% \\
\Xhline{4\arrayrulewidth}
\textbf{ASR Post PS-enabled Masking red + yellow (Only Keeping green not masked)} & 0.00\% & 0.00\% \\
\hline
\textbf{ASR Post PS-enabled Masking red (Keeping green and yellow not masked)} & 0.00\% & 0.00\% \\
\hline
\end{tabular}

\label{tab:attack_comparison}
\end{table}

Table \ref{tab:attack_comparison} summarizes our experimental results, comparing the effectiveness of these constrained attacks before and after applying our PS-enabled inference feature masking defense (Option B2). The masking in these tests was applied by masking with the neutral value of 0.5 as Alhussien et al. used min-max normalization. Initially, the constrained ZOO attack achieved an Attack Success Rate (ASR) of 19.79\%, while the constrained C\&W attack attained an ASR of 59.70\%. However, after applying our PS-enabled feature masking, both when masking red and yellow features (keeping only green features unmasked) and when masking only red features (keeping green and yellow unmasked), the ASR dropped to 0.00\% for both attacks. These results demonstrate the effectiveness of our proposed PS-enabled defense in mitigating constrained adversarial attacks on flow-based ML-NIDS. \textbf{Notably, the performance of both models remained consistent in these sets of experiments, maintaining near-perfect F1 and accuracy scores ($>$0.99) after applying our B2 defense.}

\subsection{Adversarial Training}

To further evaluate the effectiveness of our PS-enabled defenses, we performed adversarial training experiments. Table~\ref{tab:adv_train} summarizes the results of adversarial training on the UNSW-NB15 dataset using constrained ZOO and C\&W attacks. We generated 12,666 adversarial training samples from a total of 27,544 training samples. The constrained ZOO attack required approximately 40 minutes, while the constrained C\&W attack took around 44 minutes to generate these samples. After adversarial training, the Attack Success Rate (ASR) significantly decreased to 0.81\% for constrained ZOO and 3.89\% for constrained C\&W attacks.

Although adversarial training is widely regarded as one of the most effective approaches to defending against adversarial examples \cite{bai2021recent}, and as seen in Table \ref{tab:adv_train}, it reduces the ASR of the attacks, it has several drawbacks compared to our PS-enabled defenses. First, adversarial training is not universal; its effectiveness depends on incorporating specific attack types into the training process, making the model robust mainly against these types of attacks. As highlighted by Abou Khamis et al. \cite{abou2024could}, finding a general adversarial training defense for ML-based IDS remains challenging. In contrast, our PS-enabled defense is universal (attack independent), leveraging problem-space constraints inherent to NIDSs to effectively "tie the hands" of attackers without requiring retraining or attack-specific adjustments. Furthermore, our Option B defense is extremely lightweight (it has no time overhead) compared to adversarial training, which demands significant computational resources for generating adversarial samples and retraining models (as evidenced by the time metrics in Table \ref{tab:adv_train}. In summary, while adversarial training provides robust protection against specific attacks, its inefficiency and lack of universality make it less practical than our PS-enabled defense for real-world applications.

\begin{table}[t]
\notsotiny
\caption{Effect of Adversarial Training on Constrained ZOO and C\&W Attacks on UNSW-NB15 Dataset}
\centering
\begin{tabular}{|p{3cm}|p{2cm}|p{2cm}|}

\hline
\cellcolor{mygray} \textbf{Attack} & \cellcolor{mygray} \textbf{Constrained ZOO} & \cellcolor{mygray} \textbf{Constrained C\&W} \\
\hline
\hline
\textbf{Total Training Samples} & 27544 & 27544 \\
\hline
\textbf{Adversarial Training Samples}  & 12666 & 12666 \\
\hline
\textbf{Time to generate The Adversarial Training Samples (in minutes:seconds)}  & 40:23 & 44:10 \\
\hline
\textbf{Adversarial Test Samples}  & 2110 & 2110 \\
\hline 

\textbf{Successful Samples} & 17 & 82 \\
\hline
\textbf{ASR After Adversarial Training} & 0.81\% & 3.89\% \\
\hline

\end{tabular}

\label{tab:adv_train}
\end{table}

%\paragraph{drawbacks} 

\section{Related Work}

Defending ML-NIDS against evasion adversarial attacks involves three main strategies, as outlined by He et al. \cite{he2023adversarial}: (1) Parameter Protection: Techniques like Gradient Masking \cite{nayebi2017biologically} obscure model parameters to limit adversarial exploitation. (2) Adversarial Sample Detection: Methods such as those in \cite{feinman2017detecting} identify and filter adversarial inputs. (3) Robustness Optimization: Strategies like adversarial training integrate adversarial examples into the training process to enhance resilience \cite{abou2020evaluation, abou2020investigating}. 

Beyond these general strategies, researchers have explored defenses that focus on feature manipulation for security applications. For instance, Chen et al. \cite{chen2021cost} introduced cost-aware training, incorporating domain-specific costs into defense strategies, where manipulating features incurs varying levels of difficulty (e.g., asymmetric costs for increasing or decreasing a feature). Unlike their approach, our method leverages the PS metric as a generalizable measure for evaluating the perturb-ability of NIDS features in problem-space. PS enables the identification and elimination or masking of features that are inherently easy to perturb, thereby reducing the attack surface without adding overhead to the training or inference phases. Furthermore, our mechanism is completely independent of the ML model and can be applied to any model, unlike the method in \cite{chen2021cost}, which focuses specifically on tree ensemble models.

\section{Discussion and Lessons Learned}
\label{Related}

%This section presents key lessons learned from our exploration of this topic. 

%and discusses relevant related work.

\subsection{The Usual Suspects}

Through examining and analyzing research on problem-space evasion adversarial attacks \cite{han2021evaluating, hashemi2019towards, vitorino2023sok, vitorino2023towards, vitorino2022adaptative, yan2023automatic, homoliak2018improving, apruzzese2024adversarial}, we observed recurring traffic morphing techniques common across numerous attacks. These techniques consistently affect certain features, which we refer to as the \textbf{usual suspects}. The usual suspects primarily include Forward IAT, Forward Packet Length, and Forward Payload Size features. Whether our proposed PS method is adopted or not, we believe NIDS researchers should focus on these ``usual suspect'' features, as modifying them in the problem-space does not compromise the network or the malicious functionality of the adversarial flow.

%Whether our proposed PS method is used or not, we believe that NIDS researchers should pay attention to these usual suspects features as morphing these features in problem-space does not compromise the network or the malicious functionality of the adversarial flow.

%Even if ML-NIDS researchers, engineers, and architects choose not to adopt our PS method, we strongly recommend that they focus on these features, as they are frequently targeted for perturbation in problem-space adversarial attacks. 

\begin{comment}
   \subsection{Complicated Inter-Connection Features}

ct\_srv\_src, 
ct\_srv\_dst, 
ct\_dst\_ltm, 
ct\_src\_ltm, 
ct\_src\_dport\_ltm, 
ct\_dst\_sport\_ltm, 
ct\_dst\_src\_ltm.  
\end{comment}

\subsection{Five Features}

Sheatsley et al. \cite{sheatsley2022adversarial} argued that limiting the number of features an adversary can manipulate does not completely prevent adversarial attacks. Their findings demonstrated that adversarial samples could achieve a 50\% success rate by modifying just five randomly selected features. We would like to point out that our PS-enabled defenses do not aim to reduce the number of features but rather to select non-perturb-able features during the feature selection phase or mask perturb-able features \textbf{in the problem-space}. Thus, the fundamental purpose of our suggested defenses is to leave attackers with as few perturb-able features as possible (as seen in Table \ref{tab:example}, the attackers will probably have access to fewer than five perturb-able features).
Moreover, the results by Sheatsley et al. \cite{sheatsley2022adversarial} focused on targeting NIDS at the feature layer rather than the problem-space. Real-world network environments are far more complex, and the constraints modeled in their research, such as those related to protocols like TCP and UDP, are a small subset compared to the numerous limitations encountered in the actual problem-space, including side-effect features.

\subsection{Problem-space Evasion Adversarial Attacks are Already Extremely Hard for an Attacker}

Evasion adversarial attacks against ML-NIDS are highly complex and impractical for attackers \cite{shehaby2023adversarial}. ElShehaby et al. \cite{shehaby2023adversarial} identified key challenges, including limited attacker access to feature vectors, correlations between NIDS features, restricted knowledge of models and feature extraction, and the dynamic nature of modern ML-NIDS. Problem-space evasion attacks face additional hurdles, such as the Inverse Feature-Mapping Problem, which complicates the translation of feature-space perturbations into realistic network packet modifications \cite{pierazzi2020intriguing}, often causing problem-space attacks to fail to align with intended adversarial features. Moreover, some problem-space perturbations can sabotage attack efficacy; for example, adding delays as perturbations might reduce the effectiveness of DoS attacks. However, there are rare glimpses of practical evasion adversarial attacks against flow-based NIDS in research, such as the “blind” adversarial perturbations in \cite{apruzzese2024adversarial}.

%However, there are rare glimpses of practical evasion adversarial attacks against flow-based NIDS, such as “Blind” Adv. Perturbations in \cite{apruzzese2024adversarial}.

Thus, introducing a simple addition in the feature-selecting phase of ML-NIDS architecture through the PS scoring mechanism to eliminate or mask easily perturb-able features could decisively undermine these already highly impractical and complex problem-space evasion attacks. Furthermore, in real-world adversarial attacks, guessing often plays a pivotal role \cite{apruzzese2023real}. In the NIDS domain, adversaries typically lack access to the attacked model's internal information and cannot directly query the target system. As a result, they may need to predict the selected features of the target system \cite{shehaby2023adversarial}. If these selected features are unpredictable and difficult to perturb within the problem-space, attackers will be significantly constrained in crafting effective adversarial examples.

\subsection{Feature Reduction and Masking as Defenses Against Evasion Adversarial Attacks}

Feature squeezing and reduction have been explored as defenses, narrowing the attacker's search space \cite{vitorino2023sok, ibitoye2025threat}. However, our approach goes beyond simple feature reduction by applying domain-specific constraints to select resilient features using PS during feature selection (our Option A Defense). These features resist perturbations that would either compromise the network's functionality or render the attack ineffective. Furthermore, adding random padding \cite{xie2017mitigating} or noise \cite{zhang2019defending} or masking specific pixels \cite{ingle2024enhancing} has been explored as defenses against evasion adversarial attacks. However, to the best of our knowledge, our PS-enabled feature masking is the first to mask specific features based on their susceptibility to problem-space perturbations in the flow-based ML-NIDS domain. By focusing on resilient features, our defense strategies significantly reduce the effectiveness of problem-space adversarial attacks on flow-based ML-NIDS, making it much harder for attackers to succeed without disrupting network operations or their malicious objectives.

\subsection{We are losing information!}

Some may argue that utilizing PS to drop or mask certain features could result in the loss of information that may be important for the ML-NIDS. While this concern is valid, it is essential that the application of PS be guided by architects, practitioners, and engineers with substantial domain knowledge. On the other hand, our results might suggest that current literature might be utilizing more information and features than necessary. Tables \ref{tab:res1}, \ref{tab:res2A} and \ref{tab:res2} present promising results when only low-perturb features are selected or when high-perturb features are masked. Another pertinent question is: why rely on features that are easily perturb-able by an outsider?

\section{Limitations}
\label{limitations}
Although our PS-enabled defenses are equally effective in eliminating the impact of evasion adversarial attacks against ML-NIDS, they are evasion-attack-agnostic, model-agnostic, and lighter than legacy approaches such as adversarial training; however, their stability varies across implementations. Options A and B1 demonstrate robust and stable performance by integrating PS-guided feature selection or masking during training. Option B2, while the lightest defense (requiring no retraining or architectural changes) and \textbf{equally effective at neutralizing attacks}, exhibits instability in some scenarios, as seen in Table \ref{tab:performance_metrics} (e.g., SVM’s F1-score drop to 0.44 in cyan cells). We acknowledge that because masking in Option B2 is applied only during inference, it may mask features that the model relies on for accurate predictions, leading to performance degradation in certain cases.

However, post-processing techniques might mitigate these performance issues. For instance, applying class reweighting, different/dynamic masking values, probability calibration, and F1-optimized thresholding might significantly improve results for vulnerable models like SVM. We applied a combination of these mitigations, and this increased the SVM’s F1-score from 0.44 (in Table \ref{tab:performance_metrics}) to 0.96 for class 0 (benign traffic) and 0.82 for class 1 (malicious traffic), with a weighted average F1 of 0.93. By calibrating probabilities to reflect true class distributions and dynamically selecting thresholds that balance precision and recall, we restore performance without compromising the defense’s lightweight nature. Thus, while Option B2 may require supplemental post-processing for unstable edge cases (a tradeoff we plan to investigate further) it remains a viable on-the-fly defense for resource-constrained deployments in some cases.

\section{Conclusion}
\label{Conclusion}

%This paper introduces the Perturb-ability Score (PS) metric to quantify the susceptibility of features within Network Intrusion Detection Systems (NIDS) to adversarial manipulation. 
%Our research demonstrates that features with high PS are easily perturb-able in the problem-space while maintaining NIDS domain constraints, whereas features with low PS are either infeasible to alter or risk invalidating the network flow or malicious objective if modified. 
The PS metric provides a novel approach to enhancing the robustness of flow-based Machine Learning (ML)-NIDS against problem-space evasion adversarial attacks. We proposed two PS-enabled defense mechanisms: PS-guided feature selection, which utilizes only features with low PS during the feature selection phase, and PS-guided feature masking, which masks high PS features with neutral values. Both approaches leverage NIDS domain problem-space constraints to mitigate the impact of adversarial attacks on flow-based ML-NIDS. Our experimental results demonstrate that these PS-enabled defenses significantly enhance the robustness of ML-NIDS against adversarial threats while maintaining high detection accuracy. 

The PS metric and associated defenses offer several key advantages: universal applicability across various evasion adversarial attacks, ML models and datasets in the flow-based NIDS domain, lightweight implementation, practicality in addressing real-world attack scenarios, and complementarity to existing defenses. This work opens up promising avenues for future research, including refinement of the PS metric, cross-domain application, dynamic PS adaptation, and integration with advanced ML techniques. Overall, this research presents a significant advancement in the field of ML-NIDS security, providing a practical, efficient, lightweight, and effective approach to enhancing the robustness of flow-based ML-NIDS against problem-space evasion adversarial attacks.

\section*{Acknowledgement}
This work was supported by the Natural Sciences and Engineering Research Council of Canada (NSERC) through the NSERC Discovery Grant program.

\bibliographystyle{IEEEtran}
\bibliography{main}

%\section{Data Availability}

\appendices

\section{Ethics Considerations}
The datasets used in this study, MCFP, UNSW-NB15 and the improved CSE-CIC-IDS2018, are publicly available and widely used in network intrusion detection research. Additionally, the proposed Perturb-ability Score (PS) metric was developed and evaluated using synthetic experiments that do not involve human participants. 

\section{Additional Information on the constrained attacks by Alhussien et al. \cite{alhussien2024constraining}}
% and our testing of our defenses against these attacks}
\label{transWork}
\subsection{Constraints imposed by Alhussien et al. \cite{alhussien2024constraining} on adversarial attacks against NIDS}
These constraints are categorized into four types: feature mutability constraints, which identify modifiable and immutable network features; feature value constraints, which enforce valid ranges or categories for features; feature dependency constraints, which preserve semantic and statistical relationships between interdependent features; and distribution-preserving constraints, which maintain the statistical distribution of original network data. To incorporate these constraints into adversarial attack generation, they augmented the optimization objective with constraint-based penalty terms.

\subsection{Details of the attacks (Constrained ZOO and C\&W Attacks) and the Attacked Models}
To test our defenses against the constrained attacks by Alhussien et al. \cite{alhussien2024constraining} we employed two constrained adversarial attacks: the Zeroth Order Optimization (ZOO) attack and the Carlini \& Wagner (C\&W) attack. Both attacks were configured with an attack confidence of 0.5, a maximum of 80 iterations, and were executed as untargeted attacks. The constrained ZOO attack targeted a Gradient gradient-boosting classifier consisting of an ensemble model with 100 estimators, a learning rate of 1.0, a maximum depth of 5, and a random state set to 0. This model was trained on a balanced dataset. The constrained C\&W attack targeted a Neural Network model with a three-layer architecture ([256]-[256]-[128]) employing ReLU activation functions and an Adam optimizer (learning rate = 0.01), trained for 100 epochs with early stopping based on test accuracy. Table \ref{tab:attack_det} summarizes the details of the attacks and attacked models. N.B., we obtained these parameters and models from the published code of Alhussien et al. \cite{alhussien2024constraining}.

\subsection{Limitations}
While we highly appreciate the work done by Alhussien et al. \cite{alhussien2024constraining}, we observed some limitations in their testing and code implementations. However, in our opinion, the main limitation is their focus on the feature-space in their testing rather than the problem-space.
We acknowledge that their constrained attack captures many NIDS constraints, but the complexity of real-world network environments significantly exceeds the constraints modeled in the feature layer. Even in their results, after adding perturbations from their constrained attacks, 25\% of C\&W perturbed samples failed to mimic valid network traffic. Nevertheless, we utilized their work to evaluate our PS-enabled defense. As illustrated in Figure \ref{fig:Circles}, we believe that defending against constrained feature-space attacks, even if not entirely practical or valid in the problem-space, will implicitly defend against more practical problem-space attacks. In other words, even if an attack is not completely valid in the problem-space, our defense will still be effective, demonstrating the robustness of our PS-enabled approach.

\begin{table}[t]
\notsotiny
\caption{Details of the attacks (Constrained ZOO and C\&W Attacks) and the Attacked Models}
\centering
\begin{tabular}{|p{3cm}|p{2cm}|p{2cm}|}

\hline
\cellcolor{mygray} \textbf{Attack} & \cellcolor{mygray} \textbf{Constrained ZOO} & \cellcolor{mygray} \textbf{Constrained C\&W} \\
\hline
\hline
\textbf{Attack Details} & 
Attack Confidence: 0.5 \newline
Max Iterations: 80 \newline
Untargeted & 
Attack Confidence: 0.5 \newline
Max Iterations: 80 \newline
Untargeted \\
\hline
\textbf{Attacked ML Model} & Gradient Boosting Classifier & Neural Network \\
\hline
\textbf{Attacked ML Model Details} & 
Ensemble model with 100 estimators, learning rate=1.0, max\_depth=5, random\_state=0. Trained on balanced dataset. & 
3-layer architecture (input→[256][256][128] →2 outputs) with ReLU activation. Adam optimizer (lr=0.01), trained for 100 epochs with early stopping based on test accuracy. \\
\hline
\end{tabular}

\label{tab:attack_det}
\end{table}

\onecolumn

\clearpage
\section{Definition of Features of UNSW-NB15 and improved CSE-CIC-IDS2018 Datasets}
\label{AppB}
\begin{table*}[!htb]
\footnotesize
\centering
\caption{The Features Description of the UNSW-NB15 Dataset}
	\label{tab:DesUNSW}
\begin{tabular}{|l|l|}
\hline
\textbf{Feature}    & \textbf{Description}                                                                                                                  \\ \hline
srcip               & Source IP address                                                                                                                     \\ \hline
sport               & Source port number                                                                                                                    \\ \hline
dstip               & Destination IP address                                                                                                                \\ \hline
dsport              & Destination port number                                                                                                               \\ \hline
proto               & Transaction protocol                                                                                                                  \\ \hline
state &
  \begin{tabular}[c]{@{}l@{}}Indicates to the state and its dependent protocol, e.g. ACC, CLO, CON, ECO, ECR, FIN, I\\  NT, MAS, PAR, REQ, RST, TST, TXD, URH, URN, and (-) (if not used state)\end{tabular} \\ \hline
dur                 & Record total duration                                                                                                                 \\ \hline
sbytes              & Source to destination transaction bytes                                                                                               \\ \hline
dbytes              & Destination to source transaction bytes                                                                                               \\ \hline
sttl                & Source to destination time to live value                                                                                              \\ \hline
dttl                & Destination to source time to live value                                                                                              \\ \hline
sloss               & Source packets retransmitted or dropped                                                                                               \\ \hline
dloss               & Destination packets retransmitted or dropped                                                                                          \\ \hline
service             & http, ftp, smtp, ssh, dns, ftp-data ,irc  and (-) if not much used service                                                            \\ \hline
Sload               & Source bits per second                                                                                                                \\ \hline
Dload               & Destination bits per second                                                                                                           \\ \hline
Spkts               & Source to destination packet count                                                                                                    \\ \hline
Dpkts               & Destination to source packet count                                                                                                    \\ \hline
swin                & Source TCP window advertisement value                                                                                                 \\ \hline
dwin                & Destination TCP window advertisement value                                                                                            \\ \hline
stcpb               & Source TCP base sequence number                                                                                                       \\ \hline
dtcpb               & Destination TCP base sequence number                                                                                                  \\ \hline
smeansz             & Mean of the packet size transmitted by the src         %?ow                                                                
\\ \hline
dmeansz             & Mean of the packet size transmitted by the dst                                                                                    \\ \hline
trans\_depth        & Represents the pipelined depth into the connection of http request/response transaction                                               \\ \hline
res\_bdy\_len       & Actual uncompressed content size of the data transferred from the server’s http service.                                              \\ \hline
Sjit                & Source jitter (mSec)                                                                                                                  \\ \hline
Djit                & Destination jitter (mSec)                                                                                                             \\ \hline
Stime               & record start time                                                                                                                     \\ \hline
Ltime               & record last time                                                                                                                      \\ \hline
Sintpkt             & Source interpacket arrival time (mSec)                                                                                                \\ \hline
Dintpkt             & Destination interpacket arrival time (mSec)                                                                                           \\ \hline
tcprtt              & TCP connection setup round-trip time, the sum of ’synack’ and ’ackdat’.                                                               \\ \hline
synack              & TCP connection setup time, the time between the SYN and the SYN\_ACK packets.                                                         \\ \hline
ackdat              & TCP connection setup time, the time between the SYN\_ACK and the ACK packets.                                                         \\ \hline
is\_sm\_ips\_ports  & \begin{tabular}[c]{@{}l@{}}If source (1) and destination (3)IP addresses equal and port numbers (2)(4)  equal then,\\ this variable takes value 1 else 0           \end{tabular} \\ \hline
ct\_state\_ttl      & No. for each state (6) according to specific range of values for source/destination time to live (10) (11).                           \\ \hline
ct\_flw\_http\_mthd & No. of flows that has methods such as Get and Post in http service.                                                                   \\ \hline
is\_ftp\_login      & If the ftp session is accessed by user and password then 1 else 0.                                                                    \\ \hline
ct\_ftp\_cmd        & No of flows that has a command in ftp session.                                                                                        \\ \hline
ct\_srv\_src        & \begin{tabular}[c]{@{}l@{}}No. of connections that contain the same service (14) and source address (1) in\\ 100 connections according to the last time (30).       \end{tabular} \\ \hline
ct\_srv\_dst        & \begin{tabular}[c]{@{}l@{}}No. of connections that contain the same service (14) and destination address (3)\\ in 100 connections according to the last time (30).  \end{tabular} \\ \hline
ct\_dst\_ltm        & \begin{tabular}[c]{@{}l@{}}No. of connections of the same destination address (3) \\in 100 connections according to the last time (30).                             \end{tabular} \\ \hline
ct\_src\_ ltm       & \begin{tabular}[c]{@{}l@{}}No. of connections of the same source address (1) in 100\\ connections according to the last time (30).                                 \end{tabular} \\ \hline
ct\_src\_dport\_ltm &\begin{tabular}[c]{@{}l@{}} No of connections of the same source address (1) and the destination port (4)\\ in 100 connections according to the last time (30).      \end{tabular} \\ \hline
ct\_dst\_sport\_ltm &\begin{tabular}[c]{@{}l@{}} No of connections of the same destination address (3) and the source port (2)\\ in 100 connections according to the last time (30).      \end{tabular} \\ \hline
ct\_dst\_src\_ltm   & \begin{tabular}[c]{@{}l@{}}No of connections of the same source (1) and the destination (3) address \\in in 100 connections according to the last time (30).        \end{tabular} \\ \hline
\end{tabular}

\end{table*}

\begin{table*}[t]
\footnotesize
\centering
\caption{The Features Description of the CSE-CIC-IDS2018 Dataset (Part A)}
\begin{tabular}{|l|l|}
\hline
\textbf{Feature}                                   & \textbf{Description}                                                                             \\ \hline
\cellcolor[HTML]{F8F9FA}Flow ID                    & The id of the flow                                                                               \\ \hline
\cellcolor[HTML]{F8F9FA}Src IP                     & Source IP address                                                                                \\ \hline
\cellcolor[HTML]{F8F9FA}Src Port                   & Source port number                                                                               \\ \hline
\cellcolor[HTML]{F8F9FA}Dst IP                     & Destination IP address                                                                           \\ \hline
\cellcolor[HTML]{F8F9FA}Dst Port                   & Destination port number                                                                          \\ \hline
\cellcolor[HTML]{F8F9FA}Protocol                   & Transaction protocol                                                                             \\ \hline
\cellcolor[HTML]{F8F9FA}Timestamp                  & record start time                                                                                \\ \hline
\cellcolor[HTML]{F8F9FA}Flow Duration              & Flow duration                                                                                    \\ \hline
\cellcolor[HTML]{F8F9FA}Total Fwd Packet           & Total packets in the forward direction                                                           \\ \hline
\cellcolor[HTML]{F8F9FA}Total Bwd packets          & Total packets in the backward direction                                                          \\ \hline
\cellcolor[HTML]{F8F9FA}Total Length of Fwd Packet & Total size of packet in forward direction                                                        \\ \hline
\cellcolor[HTML]{F8F9FA}Total Length of Bwd Packet & Total size of packet in backward direction                                                       \\ \hline
\cellcolor[HTML]{F8F9FA}Fwd Packet Length Max      & Maximum size of packet in forward direction                                                      \\ \hline
\cellcolor[HTML]{F8F9FA}Fwd Packet Length Min      & Minimum size of packet in forward direction                                                      \\ \hline
\cellcolor[HTML]{F8F9FA}Fwd Packet Length Mean     & Average size of packet in forward direction                                                      \\ \hline
\cellcolor[HTML]{F8F9FA}Fwd Packet Length Std      & Standard deviation size of packet in forward direction                                           \\ \hline
\cellcolor[HTML]{F8F9FA}Bwd Packet Length Max      & Maximum size of packet in backward direction                                                     \\ \hline
\cellcolor[HTML]{F8F9FA}Bwd Packet Length Min      & Minimum size of packet in backward direction                                                     \\ \hline
\cellcolor[HTML]{F8F9FA}Bwd Packet Length Mean     & Mean size of packet in backward direction                                                        \\ \hline
\cellcolor[HTML]{F8F9FA}Bwd Packet Length Std      & Standard deviation size of packet in backward direction                                          \\ \hline
\cellcolor[HTML]{F8F9FA}Flow Bytes/s               & flow byte rate that is number of bytes transferred per second                                    \\ \hline
\cellcolor[HTML]{F8F9FA}Flow Packets/s             & flow packets rate that is number of packets transferred per second                               \\ \hline
\cellcolor[HTML]{F8F9FA}Flow IAT Mean              & Average time between two flows                                                                   \\ \hline
\cellcolor[HTML]{F8F9FA}Flow IAT Std               & Standard deviation time two flows                                                                \\ \hline
\cellcolor[HTML]{F8F9FA}Flow IAT Max               & Maximum time between two flows                                                                   \\ \hline
\cellcolor[HTML]{F8F9FA}Flow IAT Min               & Minimum time between two flows                                                                   \\ \hline
\cellcolor[HTML]{F8F9FA}Fwd IAT Total              & Total time between two packets sent in the forward direction                                     \\ \hline
\cellcolor[HTML]{F8F9FA}Fwd IAT Mean               & Mean time between two packets sent in the forward direction                                      \\ \hline
\cellcolor[HTML]{F8F9FA}Fwd IAT Std                & Standard deviation time between two packets sent in the forward directio                         \\ \hline
\cellcolor[HTML]{F8F9FA}Fwd IAT Max                & Maximum time between two packets sent in the forward direction                                   \\ \hline
\cellcolor[HTML]{F8F9FA}Fwd IAT Min                & Minimum time between two packets sent in the forward direction                                   \\ \hline
\cellcolor[HTML]{F8F9FA}Bwd IAT Total              & Total time between two packets sent in the backward direction                                    \\ \hline
\cellcolor[HTML]{F8F9FA}Bwd IAT Mean               & Mean time between two packets sent in the backward direction                                     \\ \hline
\cellcolor[HTML]{F8F9FA}Bwd IAT Std                & Standard deviation time between two packets sent in the backward direction                       \\ \hline
\cellcolor[HTML]{F8F9FA}Bwd IAT Max                & Maximum time between two packets sent in the backward direction                                  \\ \hline
\cellcolor[HTML]{F8F9FA}Bwd IAT Min                & Minimum time between two packets sent in the backward direction                                  \\ \hline
\cellcolor[HTML]{F8F9FA}Fwd PSH Flags              & Number of times the PSH flag was set in packets travelling in the forward direction (0 for UDP)  \\ \hline
\cellcolor[HTML]{F8F9FA}Bwd PSH Flags              & Number of times the PSH flag was set in packets travelling in the backward direction (0 for UDP) \\ \hline
\cellcolor[HTML]{F8F9FA}Fwd URG Flags              & Number of times the URG flag was set in packets travelling in the forward direction (0 for UDP)  \\ \hline
\cellcolor[HTML]{F8F9FA}Bwd URG Flags              & Number of times the URG flag was set in packets travelling in the backward direction (0 for UDP) \\ \hline
\cellcolor[HTML]{F8F9FA}Fwd RST Flags              & Number of times the RST flag was set in packets travelling in the forward direction              \\ \hline
\cellcolor[HTML]{F8F9FA}Bwd RST Flags              & Number of times the RST flag was set in packets travelling in the backward direction             \\ \hline
\cellcolor[HTML]{F8F9FA}Fwd Header Length          & Total bytes used for headers in the forward direction                                            \\ \hline
\cellcolor[HTML]{F8F9FA}Bwd Header Length          & Total bytes used for headers in the backward direction                                           \\ \hline
\cellcolor[HTML]{F8F9FA}Fwd Packets/s              & Number of forward packets per second                                                             \\ \hline
\cellcolor[HTML]{F8F9FA}Bwd Packets/s              & Number of backward packets per second                                                            \\ \hline
\rowcolor[HTML]{F8F9FA} 
Packet Length Min                                  & \cellcolor[HTML]{FFFFFF}Minimum length of a flow                                                 \\ \hline
\rowcolor[HTML]{F8F9FA} 
Packet Length Max                                  & \cellcolor[HTML]{FFFFFF}Maximum length of a flow                                                 \\ \hline
\rowcolor[HTML]{F8F9FA} 
Packet Length Mean                                 & \cellcolor[HTML]{FFFFFF}Mean length of a flow                                                    \\ \hline
\rowcolor[HTML]{F8F9FA} 
Packet Length Std                                  & \cellcolor[HTML]{FFFFFF}Standard deviation length of a flow                                      \\ \hline
\rowcolor[HTML]{F8F9FA} 
Packet Length Variance                             & \cellcolor[HTML]{FFFFFF}{\color[HTML]{333333} Minimum inter-arrival time of packet}              \\ \hline
\cellcolor[HTML]{F8F9FA}FIN Flag Count             & Number of packets with FIN                                                                       \\ \hline
\cellcolor[HTML]{F8F9FA}SYN Flag Count             & Number of packets with SYN                                                                       \\ \hline
\cellcolor[HTML]{F8F9FA}RST Flag Count             & Number of packets with RST                                                                       \\ \hline
\cellcolor[HTML]{F8F9FA}PSH Flag Count             & Number of packets with PUSH                                                                      \\ \hline
\cellcolor[HTML]{F8F9FA}ACK Flag Count             & Number of packets with ACK                                                                       \\ \hline
\cellcolor[HTML]{F8F9FA}URG Flag Count             & Number of packets with URG                                                                       \\ \hline
\cellcolor[HTML]{F8F9FA}CWR Flag Count             & Number of packets with CWE                                                                       \\ \hline
\cellcolor[HTML]{F8F9FA}ECE Flag Count             & Number of packets with ECE                                                                       \\ \hline
\cellcolor[HTML]{F8F9FA}Down/Up Ratio              & Download and upload ratio                                                                        \\ \hline
\cellcolor[HTML]{F8F9FA}Average Packet Size        & Average size of packet                                                                           \\ \hline
\end{tabular}

	\label{tab:DesCSEA}
\end{table*}

\begin{table*}[t]
\footnotesize
\centering
\caption{The Features Description of the CSE-CIC-IDS2018 Dataset (Part B)}
\begin{tabular}{|l|l|}
\hline
\textbf{Feature}                                   & \textbf{Description}                                                                             \\ \hline 
\cellcolor[HTML]{F8F9FA}Fwd Segment Size Avg       & Average size observed in the forward direction                                                   \\ \hline
\cellcolor[HTML]{F8F9FA}Bwd Segment Size Avg       & Average size observed in the backward direction                                                  \\ \hline
\cellcolor[HTML]{F8F9FA}Fwd Bytes/Bulk Avg         & Average number of bytes bulk rate in the forward direction                                       \\ \hline
\cellcolor[HTML]{F8F9FA}Fwd Packet/Bulk Avg        & Average number of packets bulk rate in the forward direction                                     \\ \hline
\cellcolor[HTML]{F8F9FA}Fwd Bulk Rate Avg          & Average number of bulk rate in the forward direction                                             \\ \hline
\cellcolor[HTML]{F8F9FA}Bwd Bytes/Bulk Avg         & Average number of bytes bulk rate in the backward direction                                      \\ \hline
\cellcolor[HTML]{F8F9FA}Bwd Packet/Bulk Avg        & Average number of packets bulk rate in the backward direction                                    \\ \hline
\cellcolor[HTML]{F8F9FA}Bwd Bulk Rate Avg          & Average number of bulk rate in the backward direction                                            \\ \hline
\cellcolor[HTML]{F8F9FA}Subflow Fwd Packets &
  \begin{tabular}[c]{@{}l@{}}The average number of packets in a sub flow in the forward direction A subflow is a TCP connection\\which can have a different internet path \\  identified by IP addresses of source and destination network interfaces.\end{tabular} \\ \hline
\cellcolor[HTML]{F8F9FA}Subflow Fwd Bytes          & The average number of bytes in a sub flow in the forward direction                               \\ \hline
\cellcolor[HTML]{F8F9FA}Subflow Bwd Packets        & The average number of packets in a sub flow in the backward direction                            \\ \hline
\cellcolor[HTML]{F8F9FA}Subflow Bwd Bytes          & The average number of bytes in a sub flow in the backward direction                              \\ \hline
\cellcolor[HTML]{F8F9FA}FWD Init Win Bytes         & Number of bytes sent in initial window in the forward direction                                  \\ \hline
\cellcolor[HTML]{F8F9FA}Bwd Init Win Bytes         & \# of bytes sent in initial window in the backward direction                                     \\ \hline
\cellcolor[HTML]{F8F9FA}Fwd Act Data Pkts          & \# of packets with at least 1 byte of TCP data payload in the forward direction                  \\ \hline
\cellcolor[HTML]{F8F9FA}Fwd Seg Size Min           & Minimum segment size observed in the forward direction                                           \\ \hline
\cellcolor[HTML]{F8F9FA}Active Mean                & Mean time a flow was active before becoming idle                                                 \\ \hline
\cellcolor[HTML]{F8F9FA}Active Std                 & Standard deviation time a flow was active before becoming idle                                   \\ \hline
\cellcolor[HTML]{F8F9FA}Active Max                 & Maximum time a flow was active before becoming idle                                              \\ \hline
\cellcolor[HTML]{F8F9FA}Active Min                 & Minimum time a flow was active before becoming idle                                              \\ \hline
\cellcolor[HTML]{F8F9FA}Idle Mean                  & Mean time a flow was idle before becoming active                                                 \\ \hline
\cellcolor[HTML]{F8F9FA}Idle Std                   & Standard deviation time a flow was idle before becoming active                                   \\ \hline
\cellcolor[HTML]{F8F9FA}Idle Max                   & Maximum time a flow was idle before becoming active                                              \\ \hline
\cellcolor[HTML]{F8F9FA}Idle Min                   & Minimum time a flow was idle before becoming active                                              \\ \hline
\cellcolor[HTML]{F8F9FA}ICMP Code                  & Internet Control Message Protocol Code                                                           \\ \hline
\cellcolor[HTML]{F8F9FA}ICMP Type                  & Internet Control Message Protocol Type                                                           \\ \hline
\cellcolor[HTML]{F8F9FA}Total TCP Flow Time        & Total TCP Flow Time                                                                              \\ \hline
\end{tabular}

	\label{tab:DesCSEB}
\end{table*}

\end{document}